\documentclass{aa}  
\usepackage{subfig,graphicx}
\usepackage{txfonts}
\graphicspath{{./figures/}}
\usepackage{natbib,twoopt} 
\usepackage[breaklinks=true]{hyperref} 
\bibpunct{(}{)}{;}{a}{}{,} 
\makeatletter
 \newcommandtwoopt{\citeads}[3][][]{\href{http://adsabs.harvard.edu/abs/#3}%
   {\def\hyper@linkstart##1##2{}%
    \let\hyper@linkend\@empty\citealp[#1][#2]{#3}}}    
 \newcommandtwoopt{\citepads}[3][][]{\href{http://adsabs.harvard.edu/abs/#3}%
   {\def\hyper@linkstart##1##2{}%
    \let\hyper@linkend\@empty\citep[#1][#2]{#3}}}      
 \newcommandtwoopt{\citetads}[3][][]{\href{http://adsabs.harvard.edu/abs/#3}%
   {\def\hyper@linkstart##1##2{}%
    \let\hyper@linkend\@empty\citet[#1][#2]{#3}}}      
 \newcommandtwoopt{\citeyearads}[3][][]%
   {\href{http://adsabs.harvard.edu/abs/#3}%
   {\def\hyper@linkstart##1##2{}%
    \let\hyper@linkend\@empty\citeyear[#1][#2]{#3}}}   
\makeatother

\begin{document} 

  \title{A VLT-ULTRACAM study of the fast optical quasi-periodic oscillations in the polar V834 Centauri
}
  \titlerunning{ QPO variabilities in V834 Cen}
   \author{
   M. Mouchet\inst{1} 
   \and J.-M. Bonnet-Bidaud   \inst{2} 
   \and L. Van Box Som \inst{3,4,2} 
   \and E. Falize \inst{2,3} 
   \and D.A.H. Buckley     \inst{5}  
   \and J.B.  Breytenbach     \inst{5}    
   \and R.P.  Ashley  \inst{6}
   \and T.R. Marsh  \inst{6}
    \and V.S. Dhillon       \inst{7,8}         
         }
    \offprints{M. Mouchet}
   \institute{
   LUTH, Observatoire de Paris, PSL Research University,  CNRS, Universit\'e Paris Diderot, Sorbonne Paris Cit\'e,  5 Place Jules Janssen, F-92195 Meudon, France (\email{martine.mouchet@obspm.fr)}
 \and  CEA Saclay, DSM/Irfu/Service d'Astrophysique, F-91191 Gif-sur-Yvette, France
 \and  CEA-DAM-DIF, F-91297 Arpajon, France
 \and LERMA, Observatoire de Paris, PSL Research Univ., CNRS, Sorbonne Universit\'es, UPMC Univ. Paris 06, F-75005, Paris, France
 \and  South African Astronomical Observatory, PO Box 9, Observatory 7935, Cape Town, South Africa
 \and  Department of Physics, University of Warwick, Coventry CV4 7AL, UK
 \and  Department of Physics and Astronomy, University of Sheffield, Sheffield S3 7RH, UK
 \and Instituto de Astrofisica de Canarias, E-38205 La Laguna,  Tenerife, Spain
} 

   \date{Received ....; accepted ...}

 
  \abstract{
   Quasi-periodic oscillations (QPOs) of a few seconds have been detected in some Polars, the synchronised 
    subclass of 
    cataclysmic systems containing a strongly magnetised white dwarf which accretes matter from a red dwarf companion. 
   The QPOs are thought to be related to instabilities of a shock formed in the accretion column, close to the white dwarf photosphere above the impact region.
   We present optical observations of the polar V834 Centauri performed with the fast ULTRACAM camera mounted on the ESO-VLT simultaneously in three filters (u', \ion{He}{ii} $\lambda 4686$, r')  to study these oscillations  and characterise their properties along the orbit when the column is seen at different viewing  angles.
   Fast Fourier transforms and wavelet analysis have been performed and the mean frequency, rms amplitude,
   and coherence of the QPOs are derived;  a detailed inspection of individual pulses has also been performed. 
   The observations confirm the probable ubiquity of the QPOs for this source at all epochs when the source is in a high state,
   with observed mean amplitude of 2.1\% (r'), 1.5\% (\ion{He}{ii}), and 0.6\% (u').
   The QPOs are present in the r' filter at all phases of the orbital cycle, with a  higher relative amplitude around the maximum
   of the light curve. 
   They are also detected in the \ion{He}{ii} and u' filters but at a lower level.  
     Trains of oscillations are clearly observed in the r' light curve and can be mimicked
by a superposition of damped sinusoids with various parameters.
The QPO energy distribution is comparable to that  of the cyclotron flux, consistent for the r' and \ion{He}{ii} 
filters but requiring a significant dilution in the u' filter. 
New 1D hydrodynamical simulations of shock instabilities, adapted to the physical 
parameters of \object{V834 Cen}, can account for 
the optical QPO amplitude and X-ray upper limit assuming a cross section of the accretion column in the range $ \sim 
(4-5)\times 10^{14}$ cm$^{2}$.  However, the predicted frequency is  larger 
than the observed one by an order of magnitude. This shortcoming indicates that the QPO generation is more complex than that produced in a homogeneous column and 
  calls for a more realistic 3D treatment of the accretion flow in  future modelling.
   }


   \maketitle

\section{Introduction}  \label{Introduction}
 Cataclysmic variables  are close interacting binaries consisting  of a white dwarf (WD) accreting matter from a red dwarf companion 
via Roche lobe overflow;   an accretion disk is generally formed around the white dwarf.
In magnetic cataclysmic variables (MCVs), the white dwarf is strongly magnetised and the accretion flow is captured from the orbital plane along the magnetic field lines  and forms an accretion column towards the WD magnetic poles.
Among MCVs, Polars are a subclass characterised by a (quasi)-synchronisation of the  white dwarf with the orbital rotation, while in Intermediate Polars, the compact object is rotating faster than the binary system. 
In Polars, no accretion disk is expected to form.
Material falling down the column is accelerated and a shock is formed near the surface of the white dwarf close to the magnetic poles.  Below the shock, the plasma cools by emitting bremsstrahlung in the X-rays and cyclotron radiation in the optical/infrared. 
These two cooling processes lead to a stratification in density and temperature. At the shock, the temperature reaches a few tens of keV, depending mainly on the mass of the
white dwarf (see reviews by 
\citeads{1990SSRv...54..195C} 
 and  
\citeads{1995CAS....28.....W}). 
Early theoretical studies 
 (\citeads{1981ApJ...245L..23L},   
  \citeads{1982ApJ...261..543C}) 
 demonstrated that the accretion is thermally unstable if it is dominated by bremsstrahlung cooling.
Shock height oscillations are expected with periods of the order of the cooling time  near the shock, typically  a few tenths of a second to seconds.
\citetads{2000SSRv...93..611W} 
presented a  comprehensive review of stationary and non-stationary models available at that date. 
Numerical simulations show that these oscillations are expected in X-rays and the optical, and that they are damped when cyclotron cooling is non-negligible 
 (\citeads{1998MNRAS.299..862S}, 
\citeads{1999PhDT........13S}, 
\citeads{2000SSRv...93..611W}, 
 \citeads{2015A&A...579A..25B}). 

Based on this cooling instability model, quasi-periodic oscillations (QPOs) were expected to be rather common in Polars. However, after an initially successful discovery time when rapid QPOs ($\sim$ 1-3 s)  were detected in the optical light of five polars, 
 no further detections were reported despite the growing number of newly identified polars,  now amounting to more than one hundred. 
The origin of these QPOs is therefore still a subject of controversy.
After the first discovery of QPOs  in the two systems V834 Cen and \object{AN UMa}
 \citepads{1982ApJ...257L..71M}, 
 three more objects were found to exhibit oscillations: \object{EF Eri}
\citepads{1987A&A...181L..15L}, 
\object{VV Pup}
\citepads{1989A&A...217..146L}, 
and \object{BL Hyi}
\citepads{1997ApJ...489..912M}, 
the last one being discovered two decades ago.

The oscillations are all found  during high states and seem a rather persistent feature for V834 Cen and AN UMa.  
The periods are in the range (0.3-1) Hz with rms amplitudes of a few per cent. From high time resolution spectroscopy, QPOs were also reported  in the \ion{He}{ii}  $\lambda 4686$ and H$_\gamma$ line fluxes of AN UMa during a high state 
\citepads{1996A&A...306..199B}. 

Oscillations have also been searched for in the X-ray band with different satellites 
(\citeads{1997MNRAS.286...77B}, 
\citeads{1999ApJ...526..435W}, 
\citeads{1999ApJ...515..404S}, 
\citeads{2000AJ....119.1930C}, 
\citeads{2002MNRAS.332..116P}, 
\citeads{2007MNRAS.379.1209R}). 
Recently, a more complete survey of a meaningful set of polars  observed with the XMM-Newton satellite also failed to detect significant X-ray QPOs
\citepads{2015A&A...579A..24B}. 
So far, the interpretation of the polar QPOs in terms of shock cooling instabilities still lacks a definitive observational confirmation. 

In this paper we present high-quality optical data of the polar V834 Cen obtained at the VLT observatory using the triple-beam high-speed CCD camera ULTRACAM, in an effort to better describe the characteristics of the optical QPOs. 
 
 The system V834 Cen  was discovered as  a bright X-ray source by 
  \citetads{1983ApJ...264..575M}, 
 and identified as a polar on the basis of a strong polarised optical and infrared flux 
  \citepads{1983MNRAS.205P...1B}. 
 Photometric data revealed an orbital  period of  101.5 min. 
 The distance is estimated at d $\sim$ 116 pc 
 \citepads{1999ASPC..157..180B} 
 and the cyclotron polar magnetic field at B $\sim$ 23 MG 
 \citepads{1992MNRAS.256..252F}. 
 The optical flux and polarisation modulations can be interpreted mainly as an effect of the beaming of the cyclotron radiation. At minimum of the optical light curve, the cyclotron radiating pole is seen nearly face on with a minimum flux corresponding to a minimum cyclotron opacity while at maximum the radiating column above the pole is seen edge on and the cyclotron opacity and flux are maximum. In the  case of V834 Cen, the geometry of the system is such that the accreting pole is not eclipsed and is always in the hemisphere of the white dwarf facing us with an orbital inclination and magnetic colatitude of $i \sim(45-50)^{\circ}$  and $\beta \sim(15-25)^{\circ}$,  respectively
(\citeads{1983MNRAS.205P...1B}, 
  \citeads{1989MNRAS.236..935C}, 
  \citeads{1990ApJ...357..582F}, 
  \citeads{2002ApJ...578..439M}, 
 \citeads{2004MNRAS.348..316P}). 
 
 After their discovery, QPOs with frequencies (0.2-1.2) Hz and amplitude (1-3)\,\% were also reported in V834 Cen by 
 \citetads{1991ApJ...382..315M}, 
 \citetads{1992A&A...265..133L},   
and
\citetads{2000PASP..112...18I}. 
The relative QPO amplitude was found to be modulated with the orbital period, with zero modulation coinciding with minimum brightness.
Different colour dependencies of the QPOs were given from non-simultaneous observations, with maximum amplitude either in the R or B band 
(\citeads{1991ApJ...382..315M}, 
\citeads{1992A&A...265..133L}).   
Upper limits for QPOs in X-rays were also reported 
(\citeads{1985ESASP.236..155B}, 
\citeads{1997MNRAS.286...77B}, 
\citeads{2000PASP..112...18I}) 
with the most stringent limit provided by a long XMM-Newton observation 
\citepads{2015A&A...579A..24B}. 

In order to test whether the QPO characteristics are compatible with a cyclotron origin, we obtained high time resolution data simultaneously at three different wavelengths, including two broadband filters in u' (3557\,\AA) and r' (6261\,\AA), and  a narrowband filter covering the \ion{He}{ii} $\lambda 4686$ line. 
The choice of the \ion{He}{ii}  emission line was motivated by the attempt to test whether QPOs might also partly come from line emitting regions as suggested by AN UMa observations 
 \citepads{1996A&A...306..199B}. 

 In this paper we  test the standard accretion model, which predicts  an oscillating shock with modulated cyclotron emission. An additional aim is to test  whether the observed  broad frequency distribution of the QPO is due to a superposition of simultaneous oscillations with different frequencies or cumulative contribution over time of individual oscillations at a given frequency. We also compare the properties of the optical oscillations with the predictions of 1D simulations in the non-stationary regime. 

In Section \ref{Observations}, the observational data are described,  and the different observational results are presented in Sect. \ref{Orbital} (light curves) and in Sect. \ref{Fast QPOs} (periodograms).  In  Sect. \ref{Discussion} we discuss the energy distribution of the oscillations and compare the observations with the predictions of 1D numerical simulations from a radiation hydrodynamic code.


\section{Observations}  \label{Observations}
V834 Cen was observed on 2005 May 16,  using the high-speed CCD camera ULTRACAM \footnote{http://www.vikdhillon.staff.shef.ac.uk/ultracam/}
 (\citeads{2007MNRAS.378..825D}). 
 The camera was mounted  as a visiting instrument on the UT3 Unit (Melipal)  of the European Southern Observatory (ESO) 8.2\,m Very Large Telescope (VLT).
ULTRACAM is a triple-beam CCD camera designed to provide three-colour imaging photometry at high temporal resolution down to 0.002\,s.
The observations  were made simultaneously in  the two filters  u' (effective wavelength 3557\,\AA) and r' (effective wavelength 6261\,\AA) of the Sloan Digital Sky Survey (SDSS) and in  a  narrowband filter  (central wavelength 4662\,\AA, FWHM 108\,\AA),  covering the emission lines \ion{He}{ii} 4686\,\AA\; and the Bowen complex \ion{C}{iii}-\ion{N}{iii}  at 4640-50\,\AA\, and designated  as the `He filter' in the following. 

The total duration of the observation is 5.58\,h, covering slightly more than three orbital cycles, with nearly continuous coverage, except for  three small gaps of 2\,s and one of 40\,s.
Data were obtained in a two-window drift mode with a 4x4 binning configuration. The exposure time was of 0.0473\,s with a dead time of 0.0039\,s, giving a resolution of 0.05122\,s. 
The exposure time was selected to give a good sampling  of the expected 2\,s quasi-periodic pulses. 
The observation times were corrected for the light travel time to the solar system  barycentre. 

The weather  was good with  typical seeing varying between 0.5 and 1 arcsec and good atmospheric transparency. 
The air mass was always lower than 1.33. 
 The data were reduced using the standard ULTRACAM pipeline software \footnote{http://deneb.astro.warwick.ac.uk/phsaap/software/ultracam/html/}, correcting for bias and flat field.
To check for variation of the sky transmission and the stability of the whole instrument, we performed aperture photometry on the target and a comparison star  in the field acquired simultaneously.
The mean level of V834 Cen was about 0.14, 0.28, and 1.60 times the level of the comparison star, respectively  in the r', He, and u' filters.
 During the first hour  of the observation, a bias problem occurred for the  r' channel and the data were rescaled using the comparison star.
 No absolute photometry was derived since we were mostly concerned with the study of rapid variability. 
 The high count rate and the typical shape of the light curve indicate that the source was in a high accretion state. 
 After correction from the extinction, the mean count rates give a reasonable estimate of the average magnitude of 15.0 and 15.8 for the r' and u' filters, respectively. 


\section{Orbital light curves}   \label{Orbital}
 
\begin{figure}
   \centering
    \includegraphics*[width=8.9cm,angle=-0,trim=80 80 40 20]{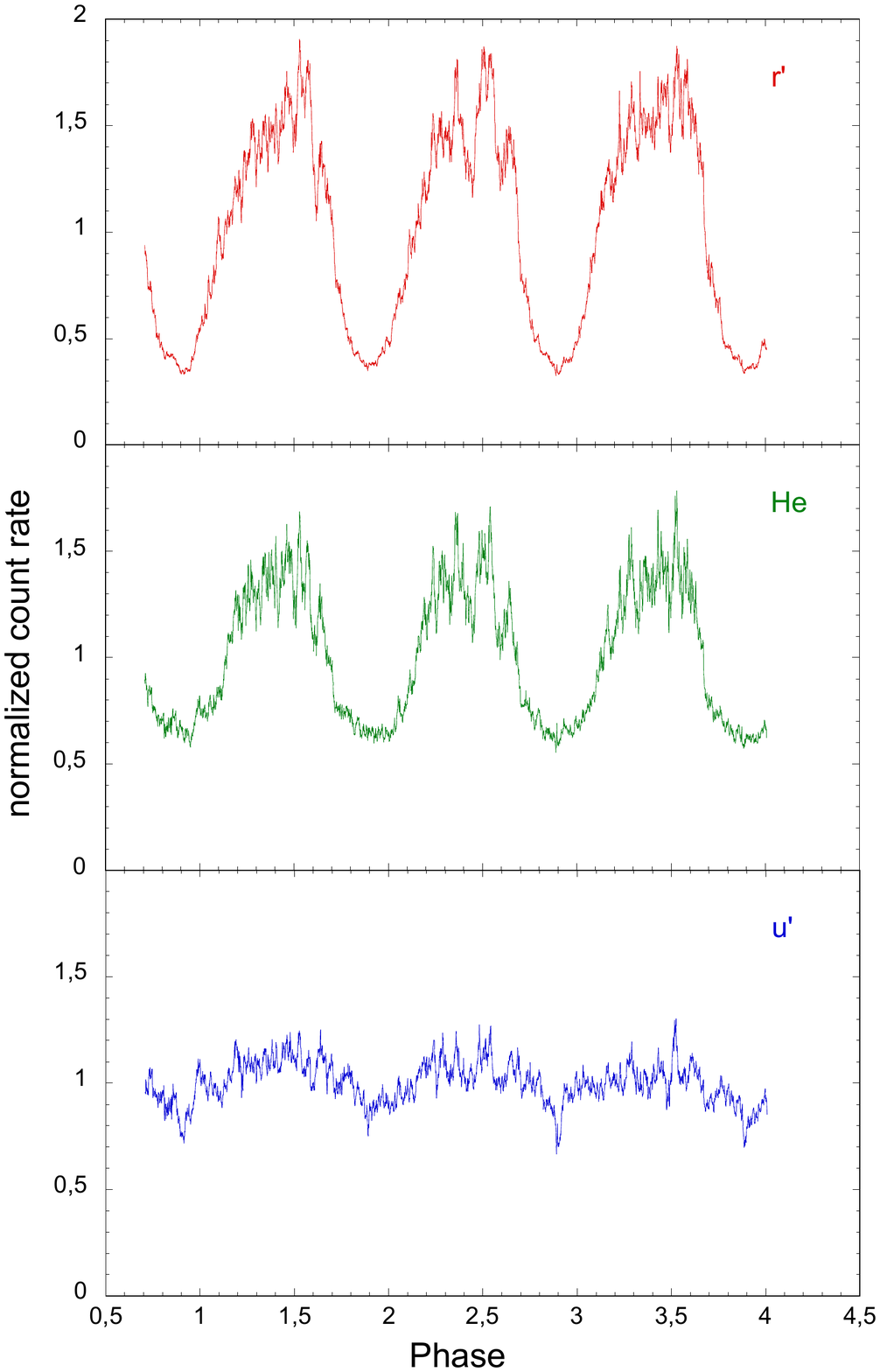}
      \caption{Normalised light curves in the r' (top), He (middle), and u' (bottom) filters versus orbital phase  at a resolution of 5.1\,s (rebin of 100 points). The continuous observation covers slightly more than three  cycles. 
                }
         \label{FigLC}
   \end{figure}

Photometric light curves were folded according to the ephemeris given by 
 \citetads{1993A&A...267..103S}, 
 where phase 0 corresponds to the blue-to-red crossing time of the secondary radial velocities.
Figure \ref{FigLC} shows the light curves for the three filters, plotted at a  5.122\,s resolution as a function of the orbital phase where the flux has been normalised to the average value.  
  
The three cycles are similar, showing a nearly sinusoidal variation with a pronounced minimum. A strong and variable flickering/flaring is also visible during the bright phase.
The r' and He light curves have a similar overall shape; the latter shows a slightly broader minimum. 
The amplitude of the orbital modulation decreases from r' to He with a peak-to-peak amplitude varying from 1.9 mag (r') to 1.1 mag (He).
The u' light curve has a different shape with a smaller $\sim$0.5 mag amplitude, a less pronounced broad minimum, and the presence of a narrow dip around $\phi \sim0.9$ superimposed on this minimum. This feature is also present, but much less visible, in the two other filters. 
Four pronounced dips are detected in u' with the last three  dips positioned at phases 0.892, 0.891, and 0.891 ($\pm 0.001$) with widths between 20 and 60\,s, while the first dip  is broader and its minimum is slightly shifted at 0.906. 
Similar deep minima are also visible in the U-band light curves shown in 
 \citetads{1983ApJ...264..575M} 
 and
 \citetads{1991ApJ...382..315M}. 
The two dip positions given in  
 \citetads{1991ApJ...382..315M} 
correspond  to a spectroscopic phase of  0.900 and 0.902 ($\pm 0.014$) given by the Schwope et al. ephemeris used  in the present paper,
thus with a consistent phase within the error bars of the present observations.
\\
For the r' and He filters, the mid-time of the broad minimum was determined by fitting a parabola to data up to half amplitude, between phase 0.75 and 1.05. The mean weighted phases of the four minima are consistently found at 0.899 $\pm$\,0.006 and 0.909 $\pm$\,0.019 respectively for the r' and He light curves. \\
We note that  observations  obtained at different dates revealed different white light curves, changing from a double-humped to a saw-tooth shape 
(\citeads{1987Ap&SS.131..643O},  
\citeads{1989MNRAS.236..935C}). 
In our observation in the r' and He filters, after a 512\,s smoothing, the second cycle appears double-humped, due to  significant increased flaring activity, while the two other cycles have a more pronounced saw-tooth shape.


\section{Fast quasi-periodic variability}   \label{Fast QPOs}
\begin{figure*}
   \centering
    \includegraphics*[width=18.2cm,angle=-0,trim=110 145 90 160]{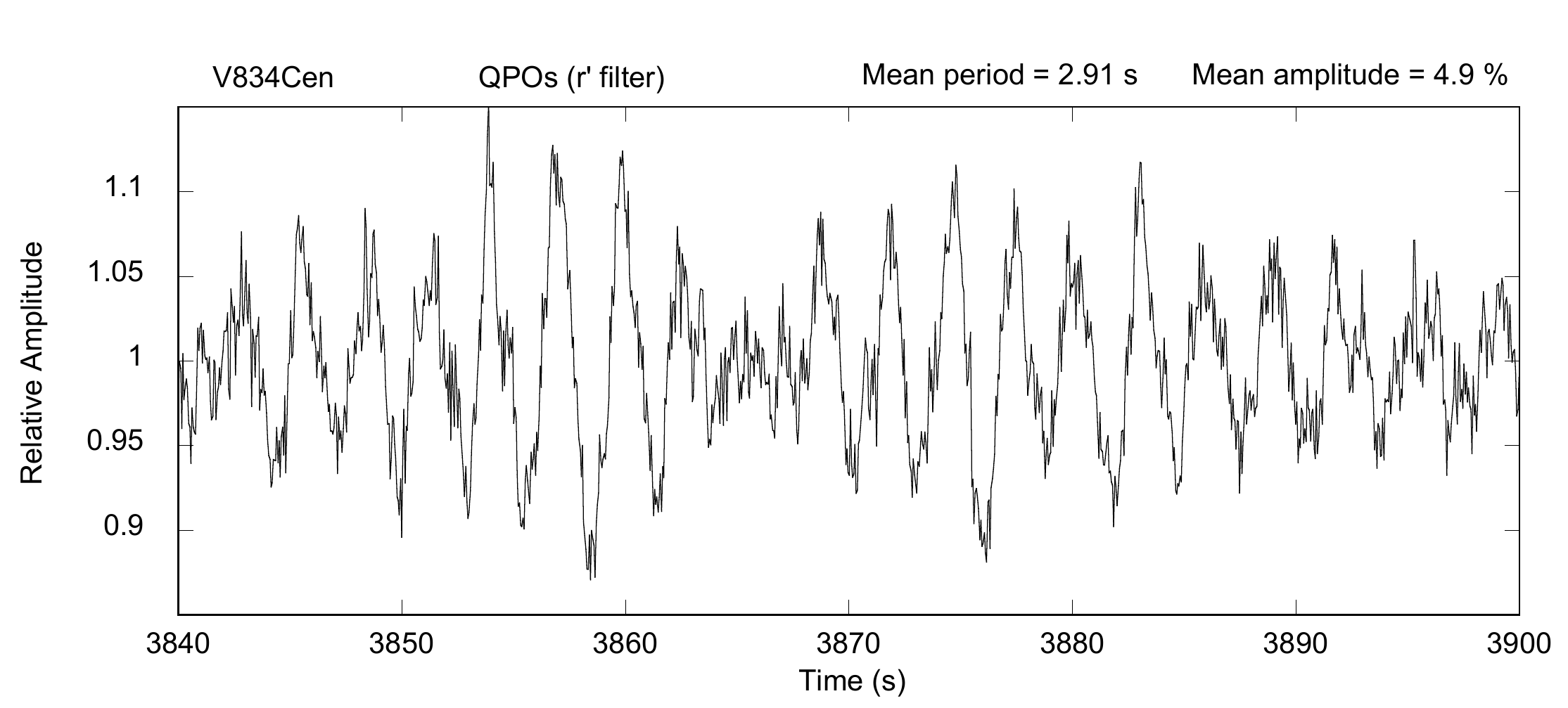}
      \caption{Typical QPOs seen in the r' light curve in a 60\,s data segment. Low-frequency variations have been removed and data normalised by dividing by a 5\,s moving average. Individual oscillation pulses are seen with variable amplitudes. A best sine fit to the data yields a mean period of 2.91\,s and mean amplitude of  4.9\%. }
          \label{bestQPOs}
   \end{figure*}

In addition to the orbital modulation and flaring variations, a close look at the light curves at high temporal resolution reveals clear variations on a timescale of a few seconds. Figure \ref{bestQPOs} shows a typical data segment of the r' light curve at the highest 0.05\,s resolution during the orbital bright phase. Quasi-periodic oscillations are directly seen displaying a sinusoidal shape with strong variable amplitudes at period around $\sim$ 2.9\,s. 
Such oscillations are also present in the brightest parts of the He light curve and are barely visible in the u' light curve.

To study the characteristics of these QPOs, fast Fourier transform (FFT) analysis of the light curves were performed for the three filters. The data set is almost continuous in time, with only four short gaps,  the longest being of 40\,s. A continuous light curve at a constant step of 0.05122\,s was built, filling the short gaps with values extrapolated from the signal immediately surrounding the gaps with typical random noise.
Standard FFTs were computed  for consecutive segments of 1024 data points  (52.45 seconds) using the {\it powspec} command of the {\it Xronos} package of the HEASOFT software\footnote{https://heasarc.gsfc.nasa.gov/docs/xanadu/xronos/xronos.html}, resulting in a total of 383 FFTs for each filter.
The obtained power  density spectra (also called FFTs in the following) extend up to 9.76 Hz (Nyquist frequency) with a bin size of 0.01906 Hz.

\subsection{Mean power spectra}   \label{Mean}
\begin{figure}
   \centering
    \includegraphics*[width=8.9cm,angle=-0,trim=60 70 70 30]{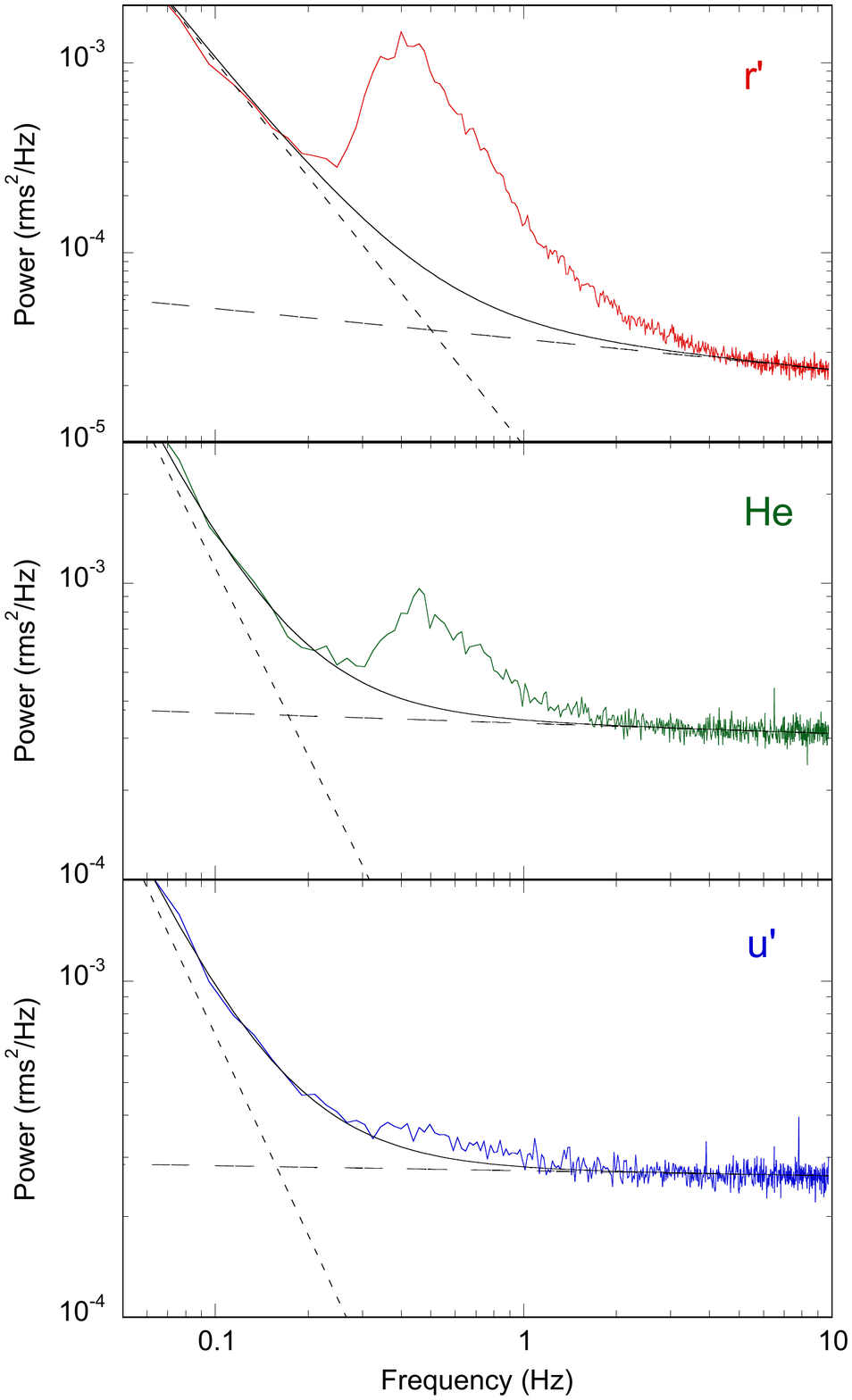}
      \caption{QPO frequency profiles. Mean power density spectra of the three  light curves  (top: r', middle: He, bottom: u') for the full $\sim$ 5.6\,hr observation are shown in logarithmic scale in the range 0.03-9.76\,Hz. The Y scale is the squared fractional amplitude in units of rms$^{2}$.Hz$^{-1}$.
      Two best fit power laws are shown representing the additional low-frequency red noise (dotted line) and high-frequency white noise (dashed line);  their sum (full line) has been subtracted to compute the QPO power (see text).
   }
        \label{FigmeanPSP}
   \end{figure}

To define the overall properties of the fast variability, mean quadratic-summed FFTs over the full observation were first computed for the three filters and are shown in 
Fig.\ref{FigmeanPSP}  in logarithmic scale. 
A broad ($\sim 0.25- 2$ Hz) bump, signature of the quasi-periodic oscillations is clearly detected for the three filters with decreasing amplitudes from r' to u'.
In addition, a significant  power is seen at low frequencies (red noise due to flickering/flaring) as well as typical  white noise at high frequencies. 
When the count rate   statistics are better, as in the case of the r' filter, a significant QPO excess over white noise is visible up to $\sim$ 5\,Hz. 
%
\begin{table}
\caption{Mean properties of the QPOs}             
\label{table:1}      
\begin{flushleft}
\begin{tabular}{ llll }        
\hline\hline                 
 & r' &He & u' \\
 \hline
 Amplitude (rms \%)    & 2.09 (0.01)  & 1.48 (0.07)  & 0.55 (0.17) \\
 Mid-freq.  $\nu_m$ (Hz)* & 0.52 (0.01)  & 0.61 (0.03) & 0.67 (0.20) \\
 FWHM (Hz) & 0.25 (0.03) & 0.26 (0.03) & 0.26 (0.07)\\
\hline                        
\hline                                   
\end{tabular}
\end{flushleft}
(*) see text for definition
\end{table}
%

To compute the mean amplitudes of the oscillations, we first subtracted best fit power laws defined at low frequencies in the range (0.03-0.2) Hz (red noise) and high frequencies in the range (5-9) Hz  (white noise).  These fits are shown in Fig.\ref{FigmeanPSP}.
The amplitudes of the QPOs were then evaluated by computing the integrated power in the range from 0.25 to 5\,Hz. Confidence intervals were computed by propagating the uncertainty in the best fit power law parameters. 
Results for the three filters are listed in Table~\ref{table:1}.
QPOs are dominant in the r' filter reaching an amplitude of 2.09\%, and show a decreasing value for the He filter (1.48\%) and u' filter (0.55\%).

As a check,  we have also evaluated the QPO amplitudes with a more straightforward measurement by detrending at low frequency using  a third-order polynomial and subtracting a constant background evaluated between 5 and 9\,Hz at high frequency. The power spectrum was then integrated between 0.25 and 5\,Hz.  Confidence intervals were obtained by varying the subtracted  background by plus or minus three times the standard deviation on the mean background value. The corresponding amplitudes for the r', He, and u' filters are respectively 2.19$\pm$0.01\%, 1.72$\pm$0.06\%, and 0.83$\pm$0.11\%. The results are comparable with the values given in Table~\ref{table:1}, but as expected, this background subtracting method slightly overestimates the QPO amplitudes. 

 The broad excess  of the QPOs, in log scale,  in the power density spectrum is clearly asymmetric so that the mean QPO frequency can be defined in different ways. Three types of measurements can be compared: 
 the frequency $\nu_p$ of the highest peak,
 the mean frequency $\nu_w$ weighted by power values (statistical mean), and
 the mid-frequency $\nu_m$ which gives  equal integrated powers in  the two ranges $\nu_{low}$-$\nu_m$  and $\nu_m$-$\nu_{high}$ (median).
 The computations were performed on the power spectrum after red noise and white noise subtraction and in the selected frequency range (0.25-5) Hz. 
 
 The mid-frequency $\nu_m$ is listed in Table 1 with values 0.52, 0.61, and 0.67 Hz respectively for the  r', He, and u' filters.
 The asymmetry of the QPO bump with a positive skewness provides lower values for $\nu_p$, with respectively 0.40, 0.46, and 0.50 Hz for the  r', He, and u' filters; instead, the weighted frequency $\nu_w$ yields higher values at 0.68 and 0.76 for the r' and He filters and gives unreliable results for the u' filter, due to low signal-to-noise ratio. 
 The typical frequency values are all in the range 0.4-0.8\,Hz. 
 Regardless of the method, there is a tendency for slightly smaller values in the r' filter than in the He filter,  but as the error bars are computed here from pure statistics, the difference may not be significant because of systematic uncertainties.
 
The QPO widths -- defined as full width at half maximum (FWHM) -- were measured for the power spectra slightly smoothed by a rebinning of 0.04\,Hz. A typical width of 0.26\,Hz is found for the three filters (Table~\ref{table:1}). The corresponding quality factor $Q$ is therefore only of the order of  2 when averaged over the full $\sim5.5$hr observation.


\subsection{Periodogram temporal variations}
\begin{figure}[h]
   \centering
  \includegraphics*[width=8.9cm,height=9.9cm,angle=-0,trim=0 30 20 90]{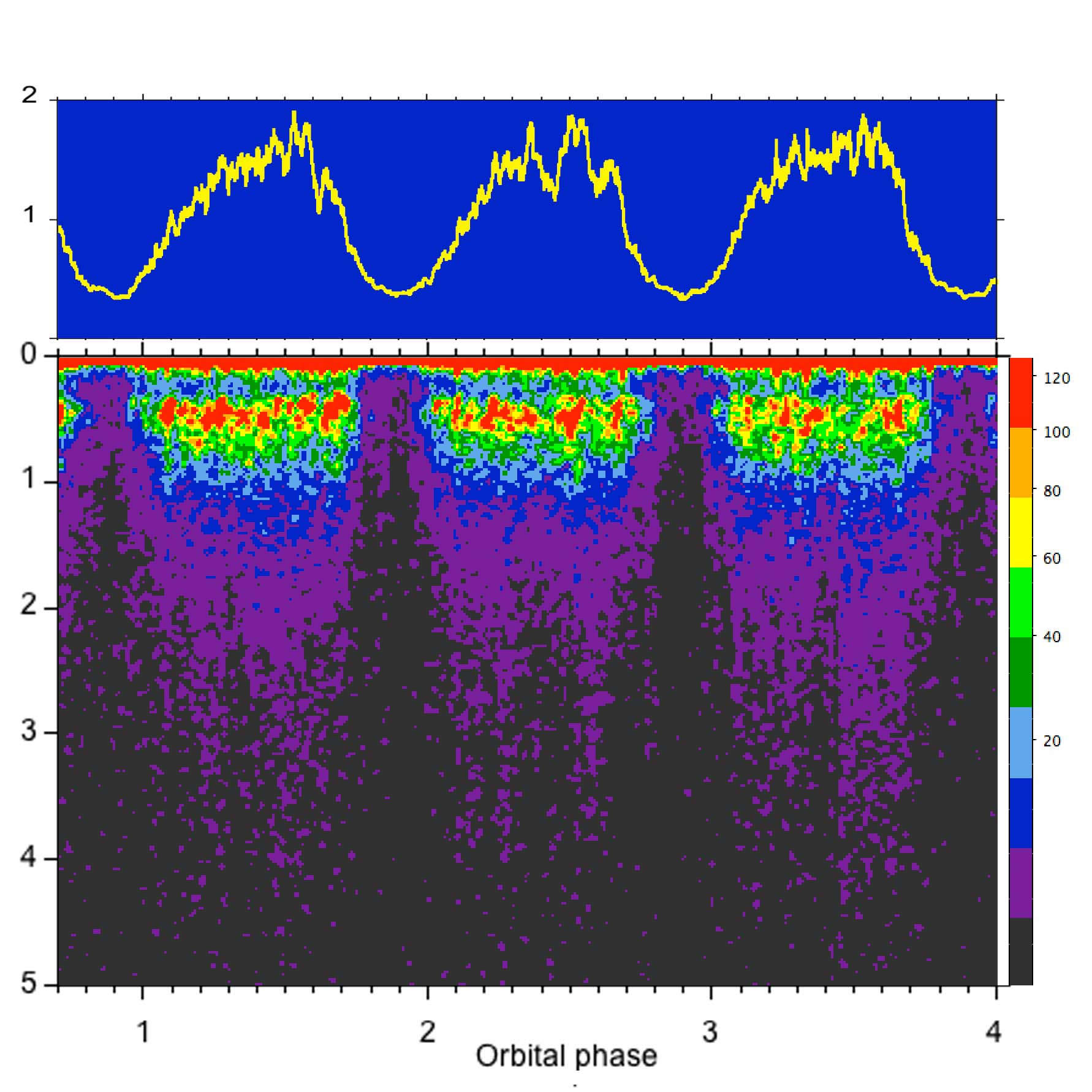}
    \caption{Two-dimensional time-frequency image of the  fractional  rms amplitude for the r' light curve. The X-axis is the orbital phase and the Y-axis is the frequency in the restricted range 0-5\,Hz. Colour-coded squared amplitude is shown according to the scale at the right (in units of $1.4\times 10^{-5}$ rms$^{2}$.Hz$^{-1}$). The individual (52.4s) spectra have been normalised to their average level in the range 5-9 Hz and the image has been smoothed with a two-point Gaussian filter. The corresponding normalised r' light curve is shown at the top.
}
          \label{Fig2Dn}
   \end{figure}

Inspection of the  52.4\,s segment light curves and their associated FFTs reveals strong time variability  for the three filters. 
To visualise this temporal variability, we built 2D time-frequency images of these FFTs after normalising the power value to the white noise level evaluated in the range 5-9 Hz. \\
The image that shows the QPO power density (relative amplitude squared) for the r' filter is displayed  in Fig. \ref{Fig2Dn} in the range [0.019-5]\,Hz. The image, built from 383 power spectra with no overlap, has been slightly smoothed by a  two-point  Gaussian function.
The QPOs are clearly seen around 0.5\,Hz during bright phases. The relative amplitude is strongly modulated with the source flux with a deep minimum centred around phase 0.9, corresponding to the light curve broad minimum. At bright phases, a significant power is seen up to $\sim$4-5 Hz. The maximum power is concentrated around $\sim$ 0.5\,Hz, showing trailing structures corresponding to significant changes in amplitude and frequency over minute timescales.
The 2D image for the He filter shows a  comparable shape but with a lower amplitude. A coherence 2D spectrum produced with the r' and He light curves indicates a high degree of coherence between these two bands.

   \begin{figure}[h]
   \centering
  \includegraphics*[width=8.9cm,height=7.5cm,angle=-0,trim=10 20 10 40]{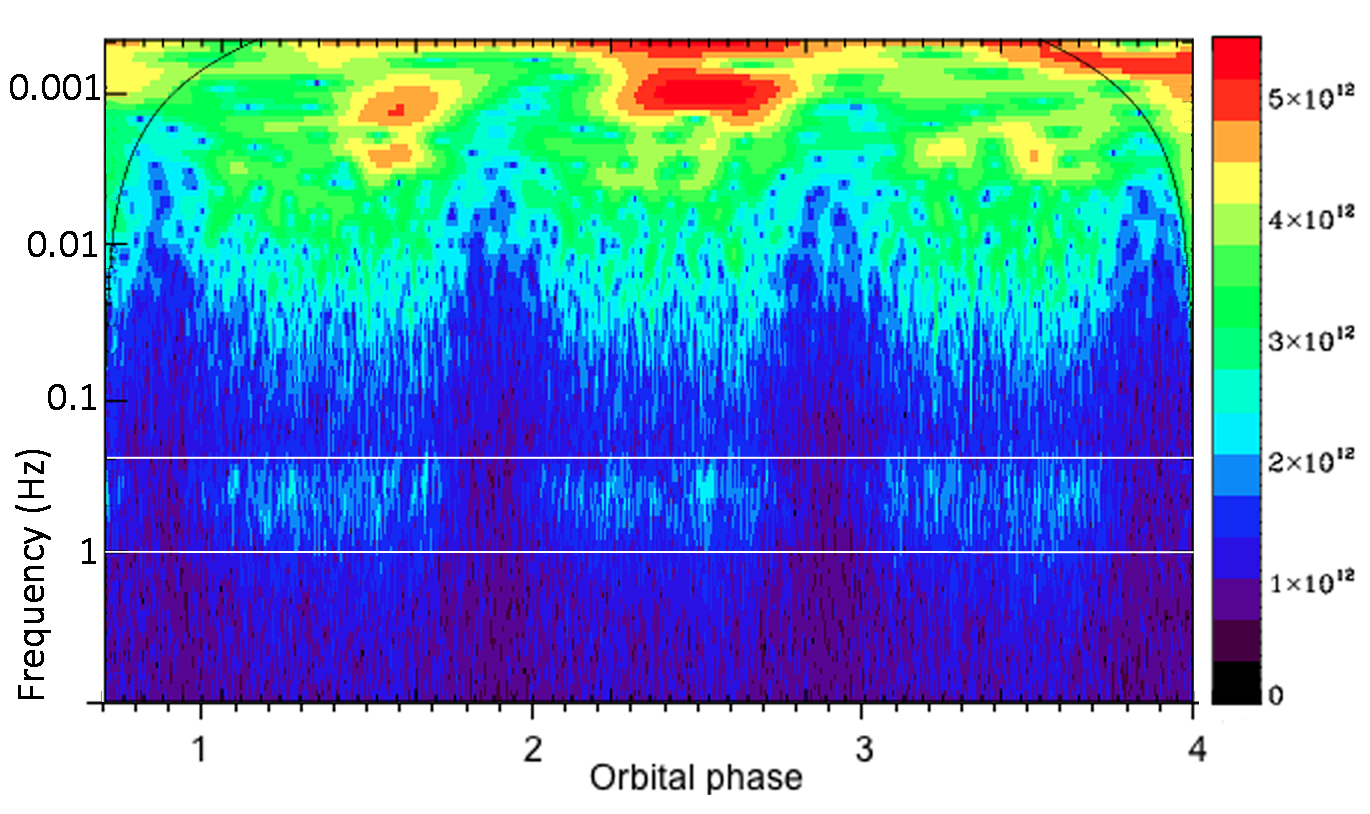} 
    \caption{Wavelet transform (scalogram) of the r' filter light curve, mapping the time-frequency evolution of the signal. Significant QPO amplitude is visible around $\sim$ 2 s (delimitated by horizontal white lines) but variability is also present in the range $\sim$100\,s. High flaring activity is also visible around 1000 s in the second orbital cycle. The thin black line at the top of the figure delimits the cone of influence below which the information is fully usable. }
          \label{Fig2Dwt}
   \end{figure}

As  significant red noise is present due to the source ($\sim$100 s) flaring activity, we also investigate the source variability using  IDL routines for wavelet analysis provided and described by 
 \citetads{1998BAMS...79...61T}. 
  This allows us to efficiently examine variability at various timescales.
We use the Morlet Gaussian wavelet basis function: a sinusoid at frequency $\nu$, modulated in amplitude by a Gaussian envelope with an extent $\tau_d$
such that the product $\tau_d \cdot \nu$ is constant.  The modulation is thus adjusted to the frequency and  allows us to scan short duration signals as well as longer ones.\\
 The resulting wavelet 2D map for the r' filter is shown in Fig. \ref{Fig2Dwt} where the signal squared amplitude is shown as a function of frequency.  
Significant amplitudes are detected around periods of $\sim$ 2 s corresponding to the QPOs, and also in the wide range 50-1000 s.  The thin black line represents the cone of influence that corresponds to the maximum period for which significant information can be derived at a given time. Periods greater than this limit are subject to edge effects (see 
\citealt{1998BAMS...79...61T}). 

\begin{figure}
   \centering
    \includegraphics*[width=9.9cm,angle=-0,trim=80 60 30 40]{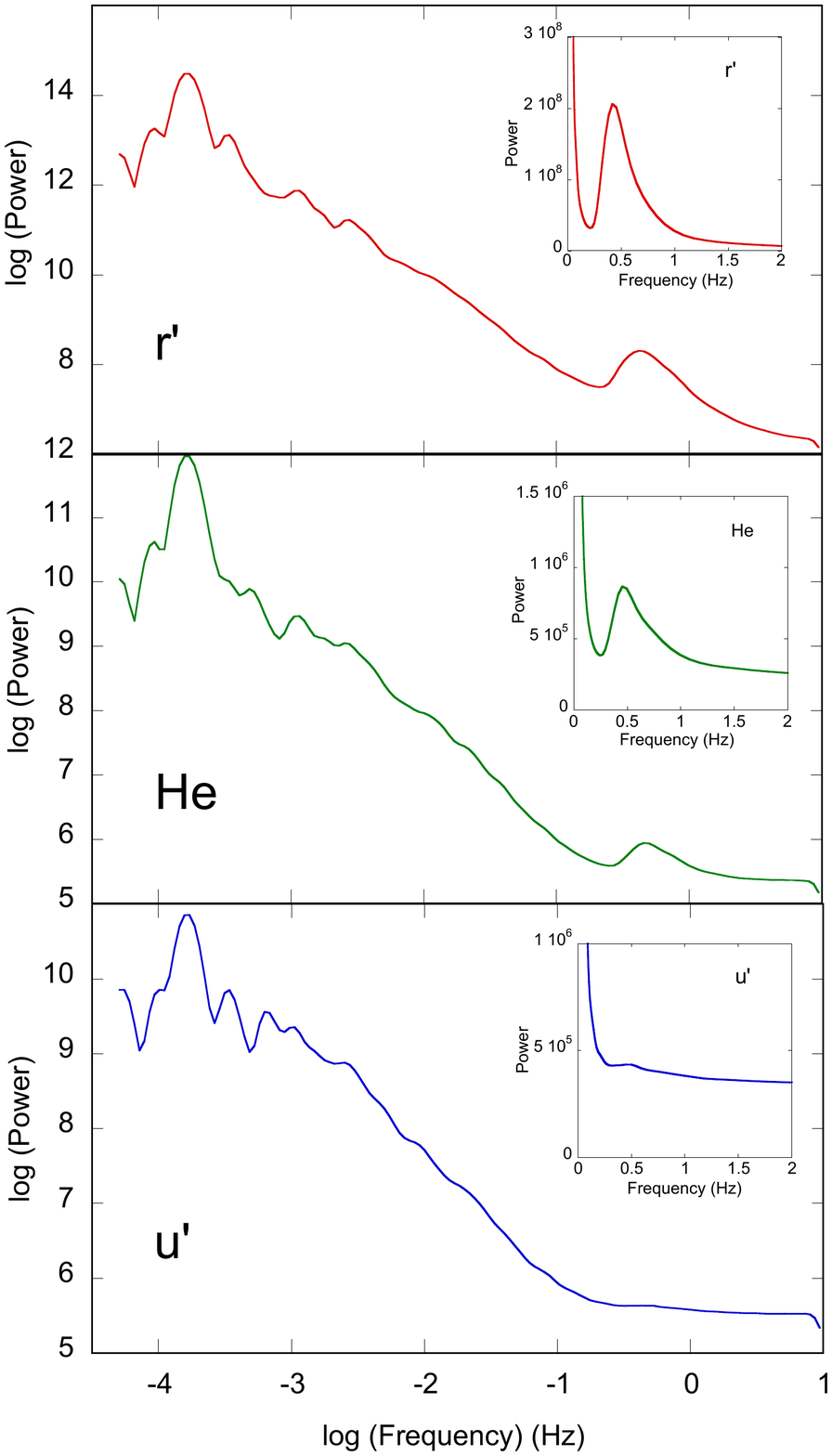}
      \caption{Main frame: Mean wavelet spectra of the light curves for the three filters (top: r', middle: He, bottom: u') shown in logarithmic scale in the range (5$\times 10^{-5}$ - 10)\,Hz. The peak at  $1.6 \times 10^{-4}$ Hz corresponds to the orbital frequency and some power is also visible in the range $(10^{-3}$-$10^{-2})$  Hz.
  Insets: Subset in the range (0-2)\,Hz is shown in linear scale to emphasise the fast QPOs.  }
        \label{Figmeanwave}
   \end{figure}

For comparison with the mean power spectrum obtained from the FFT analysis (see Sect. \ref{Mean}), we also built a 1D wavelet transform by averaging the wavelet squared amplitudes over the whole observation. The results are shown in Fig. \ref{Figmeanwave} for the three filters r', He, and u'. The full ($5\times 10^{-5}$-10 Hz) frequency range is covered in logarithmic scale in Fig. \ref{Figmeanwave} (main frame), revealing the fast QPOs as a bump with a $\sim$ 2\,s (0.5\,Hz) timescale and a strong power at the orbital period at 101.5 min ($1.64\times 10^{-4}$ Hz).  A smaller but significant excess   is also clearly visible in the 
intermediate range of minutes.
The insets in Fig. \ref{Figmeanwave} also show the amplitude in the subset (0-2\,Hz) frequency range in the more conventional linear representation, revealing the broad QPO excess with a decreasing amplitude in the three filters.
  
\subsection{QPO orbital variability}
To investigate the variations of the QPO characteristics along the orbital cycle, FFTs computed over 52.4\,s time intervals were phased and averaged in 20 orbital phase bins with each phase bin including typically 19 FFTs.
Quantitative QPO measurements were derived for the 20 phase bins in the three filters. As the signal statistics are lower than for the orbital averaged spectra, the fractional rms amplitude was computed here by subtracting a constant background at high frequency (5-9\,Hz), as described in Sect. \ref{Mean}.  The power spectrum was integrated between 0.25 and 5\,Hz.  As shown   in Sect. \ref{Mean}, this gives amplitudes nearly consistent with the more complete red and white noise corrections.
The QPO characteristic frequency was also determined by evaluating the mid-frequency, as defined in Sect. \ref{Mean}.\\

   \begin{figure}
   \centering
 \includegraphics[width=8.9cm,angle=0,trim=90 290 90 240]{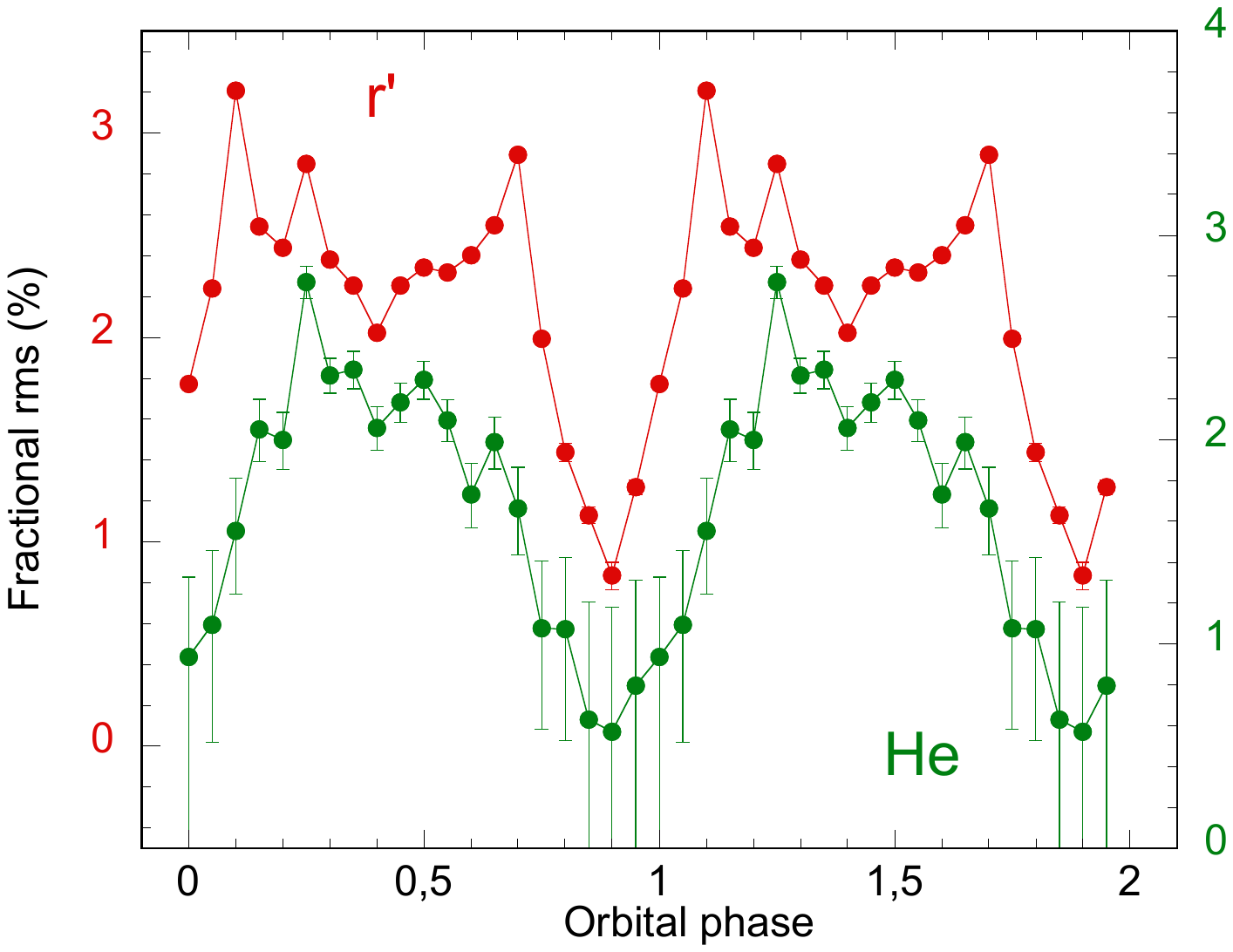}  
      \caption{Fractional rms amplitude variation as a function of the orbital phase for the r' (red dots) and He (green dots) filters. The different right (r') and left (He) scales have the same amplitude, but are shifted by 0.5 to avoid overlap.
       }
         \label{Figamplbin}
   \end{figure}

The phase averaged QPO amplitudes are shown in Fig. \ref{Figamplbin}.
Given the different filters, QPOs could not always be detected.  
In particular, no significant QPOs were detected for the u' filter in most orbital bins except one at phase 0.85, and so this filter is not shown. In the He filter, amplitudes in the phase interval (0.85-1.00) are not statistically significant.

The QPO fractional rms amplitudes measured for the r' and He filter (see Fig. \ref{Figamplbin}) are not constant along the orbit, but roughly behave as the light curves with a broad minimum around phase 0.9. The highest values are measured for both filters in the rising (phase $\sim$0.1-0.2) and the decreasing (phase $\sim$0.7) parts of the light curve with no obvious correlation with the flaring activity. The width of the broad minimum is also slightly larger in He than in r' filter, similar in shape to the corresponding light curve (see Fig. \ref{FigLC}). 
 
\begin{figure}
   \centering
     \includegraphics*[width=8.9cm,angle=-0,trim=50 190 200 140]{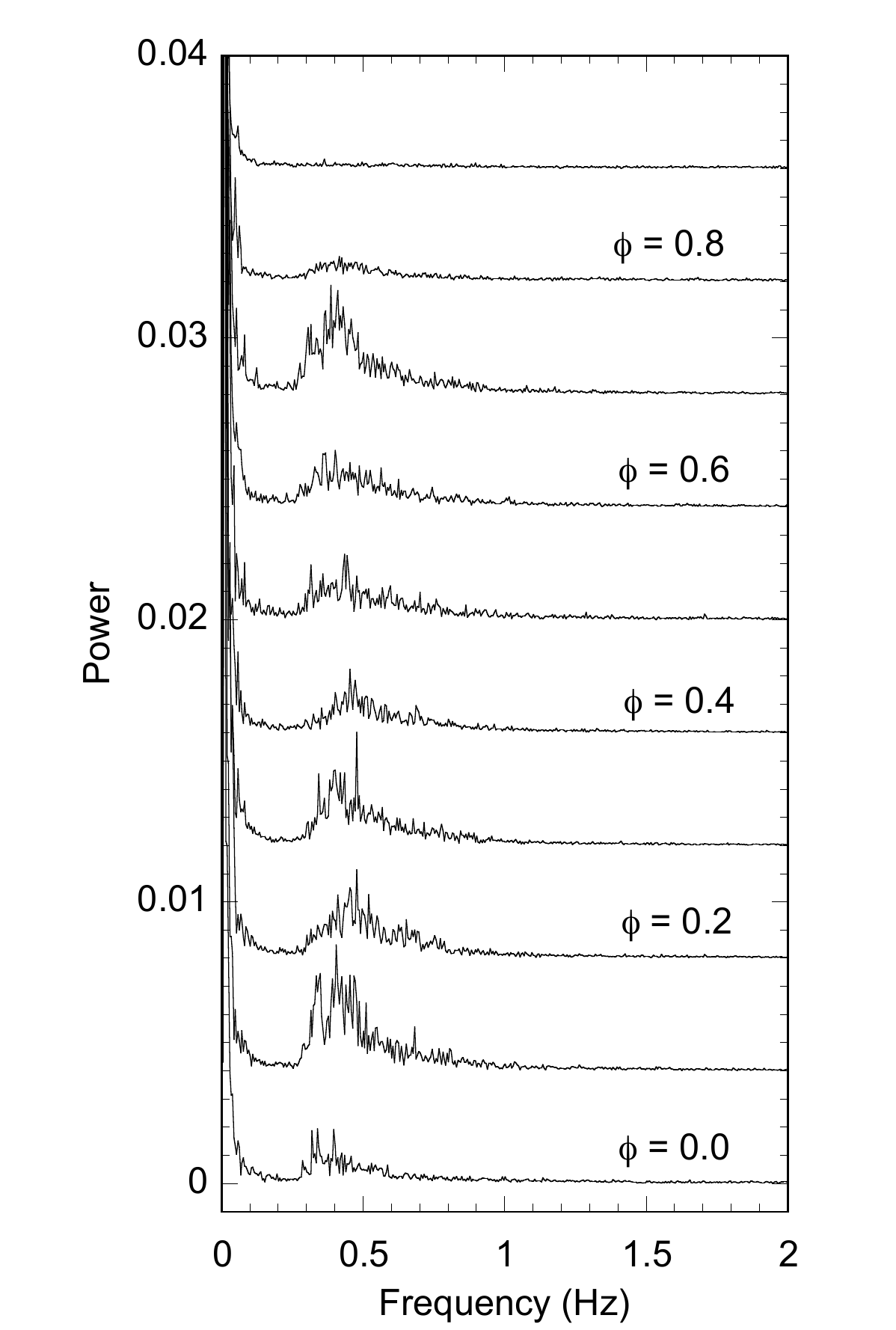}
\caption{Phase-resolved  power spectra for the r' light curve in the range (0-2)\,Hz, built from individual 210\,s FFTs. Phase  increases from bottom to top by steps of 0.1 in phase. Power is in units of rms$^{2}$.Hz$^{-1}$. Each phase spectrum is shifted vertically by a constant value of $0.004$. } 
         \label{Fig20bin}
   \end{figure}

To illustrate the orbital variability of the power density spectra  in the r' filter,  in Fig. \ref{Fig20bin} we show  power spectra in ten phase bins, built by averaging 4096-point FFTs corresponding to time duration of 210\,s, for a better frequency resolution.
The overall QPO profiles are broad, but even after averaging several individual narrow peaks are still detected in most phase bins. For r', QPOs are strongest around $\phi \sim$ 0.1 and 0.7, and are not distinguishable at $\phi \sim$ 0.9 at this scale.  \\

\subsection{QPO short-term variability}
Inspection of Fig.\,\ref{FigLC} and Fig. \ref{Fig2Dn} shows that the source behaviour is not totally reproducible from cycle to cycle; for instance,  the second cycle  shows increased flaring activity, and enhanced QPOs are also seen at different phases from cycle to cycle. 
We therefore analysed separately the 383  FFTs of 52.4\,s duration obtained using the previously described background subtracting method.

\begin{figure}
   \centering
    \includegraphics*[width=8.5cm,angle=-0, trim= 100 250 60 30]{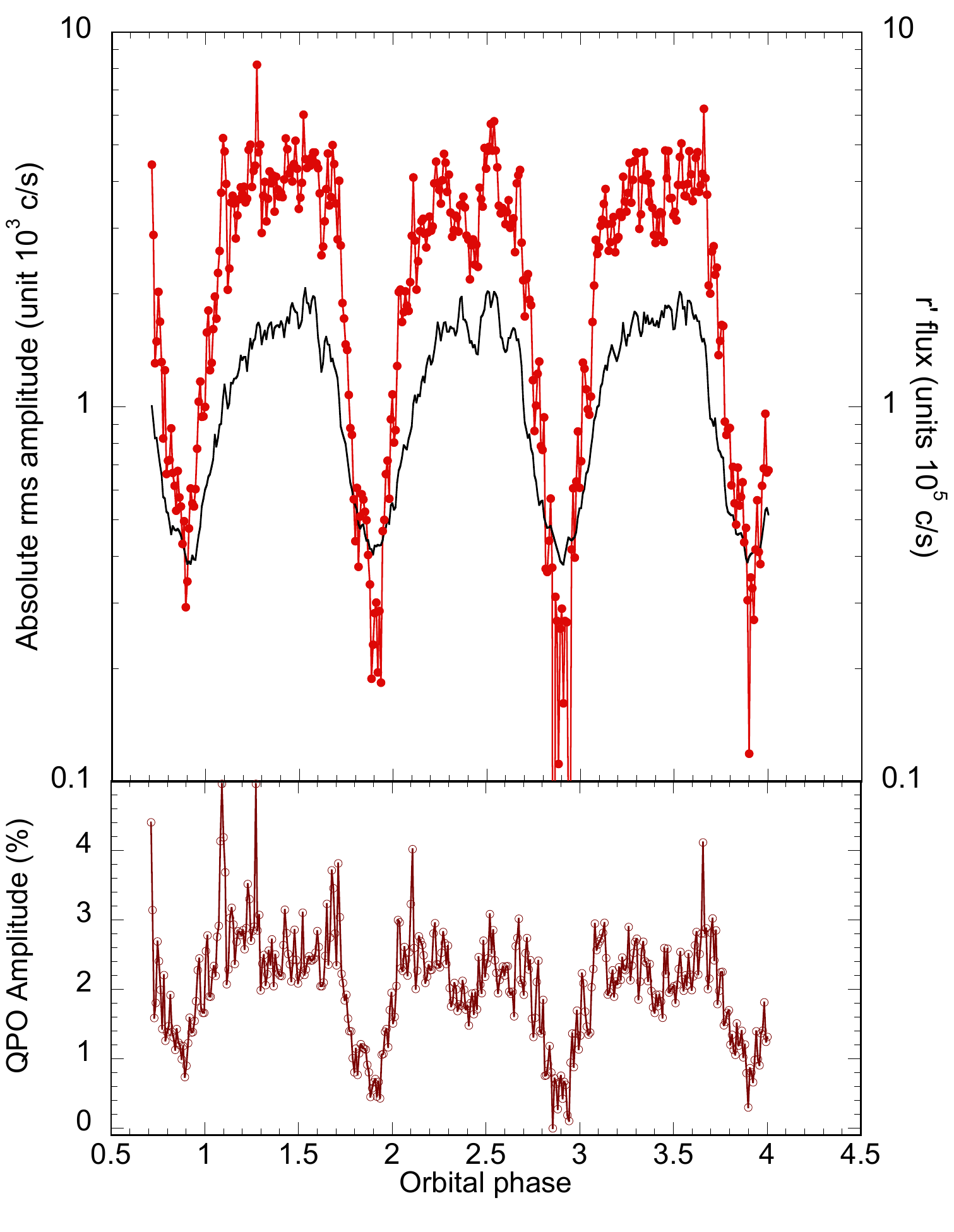}
       \caption{
     Top:  Absolute QPO rms amplitude versus orbital phase computed for the 52.4 s data segments in the r' filter, shown in logarithmic scale (in red). The r' light curve (in black) is also superposed, with the scale given at right.   Bottom: Fractional QPO rms amplitude (in per cent)  versus orbital phase.   }
         \label{Figampfluxphase}
   \end{figure}

Figure\,\ref{Figampfluxphase} (top) shows the absolute amplitudes of the QPOs in the r' filter along the observation. For comparison, the corresponding  r' light curve is also shown at the same 52.4\,s resolution. The data are plotted in logarithmic scale to show  the ratio more clearly.
At first sight, the absolute amplitude of the QPOs  seems to closely follow the overall light curve variations, reaching maxima values around $\sim 4 \times 10^{3}$ count/s during the source bright orbital phases, corresponding to a mean ratio $\sim$ 0.025  of the source flux. 
Some of the flaring activity visible in the second cycle ($\phi \sim$ 2.5)  also has a counterpart with corresponding higher QPO amplitudes. 
However, the fractional rms amplitude is  not strictly constant (Fig.\,\ref{Figampfluxphase}, bottom).  
 At low level, the fractional amplitude steadily increases with flux, before reaching a plateau  around 2.4\%,  where  systematic correlation is no longer observed.
 
The QPO amplitudes for the r' filter clearly show that significant oscillations are present during the whole observation, including the faint phase. 
At minimum flux, around $\phi \sim$ 0.9, the amplitude reaches small but still marginally significant values (see Fig. \ref{Figamplbin} and  Fig. \ref{Figampfluxphase}). 
At this phase, the oscillations are very marginally visible in the He light curve and essentially not seen in the u' light curve, due to lower count rate (weak signal-to-noise ratio). 
We note that the mean source count rates for the He and u' filters are similar and about 10 times smaller than that of r'.


    \begin{figure*}
     \centering
   \includegraphics*[width=6.0cm,height=8.5cm,angle=-0,trim=60 60 80 30]{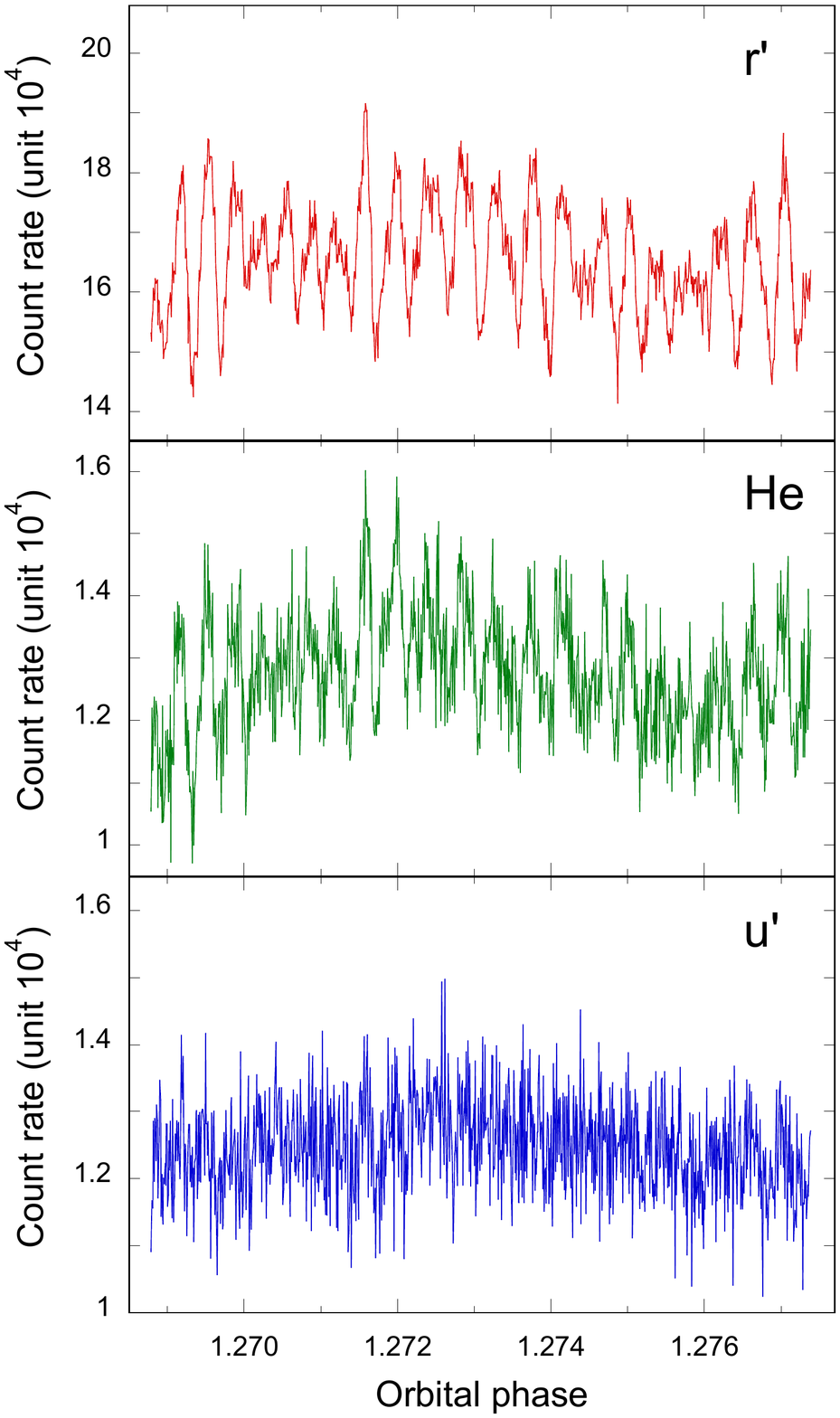}
   \hspace*{0.cm} 
     \includegraphics*[width=6.0cm,height=8.5cm,angle=-0,trim=60 60 80 30]{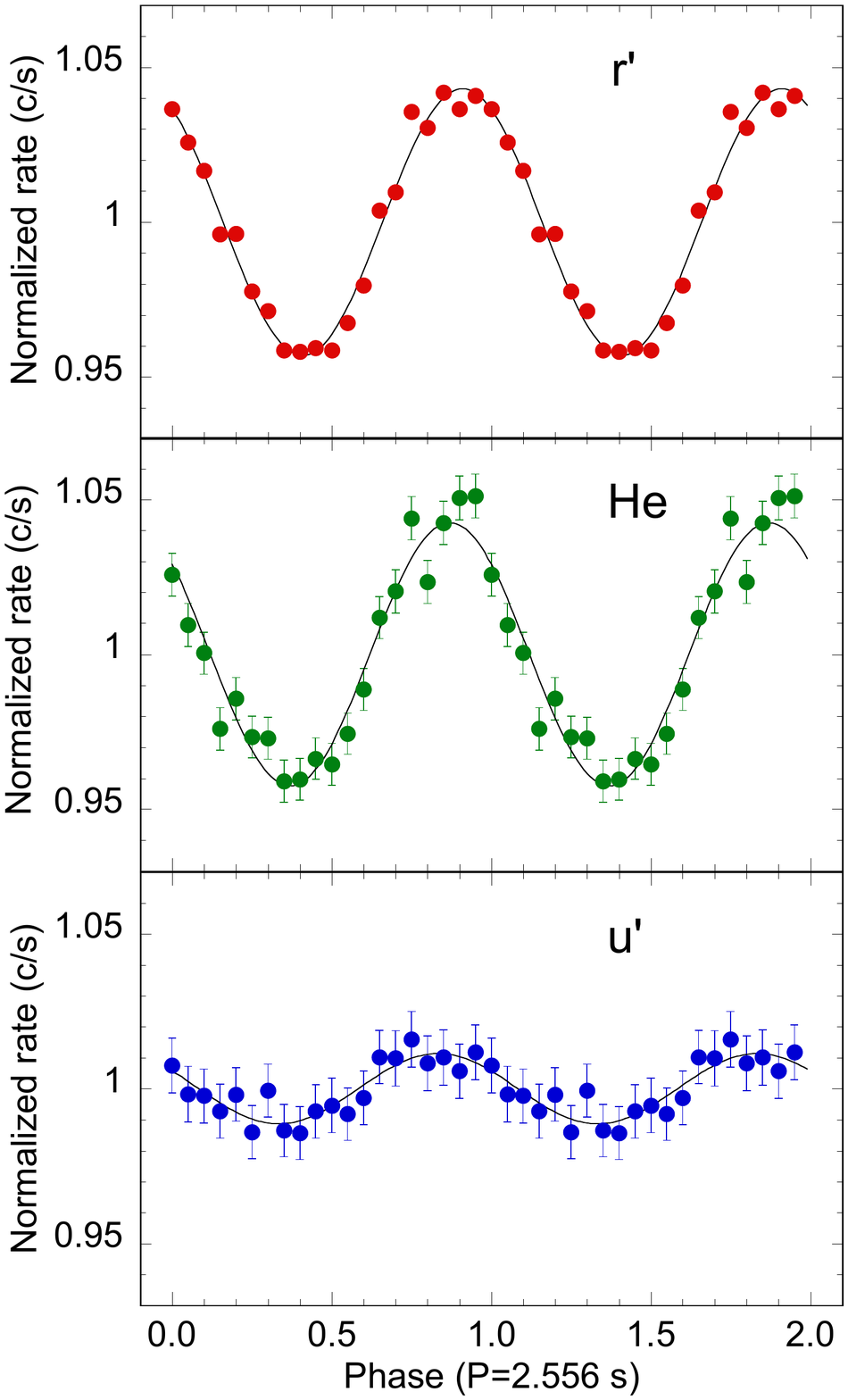}
       \hspace*{0.cm} 
   \includegraphics*[width=6.0cm,height=8.5cm,angle=-0,trim=50 60 80 30]{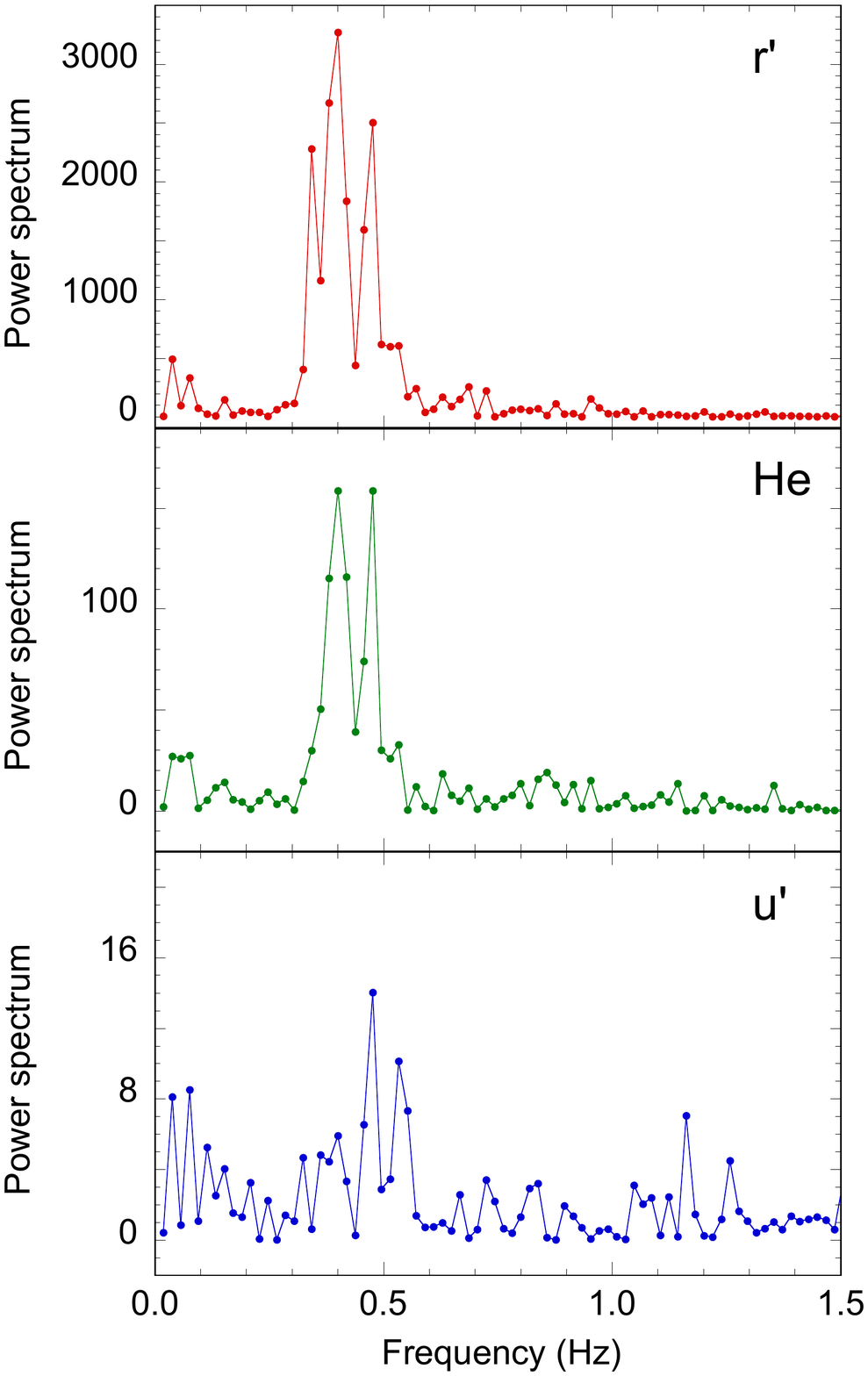}
      \caption{Left: Example of individual QPOs in a 52.4 s segment of the light curves around  $\phi=1.273$,  for r'(top, red), He (middle, green), u' (bottom, blue). 
      Middle: Pulse profile folded with the best sinusoidal fit period of 2.556\,s; pulses are reported twice for clarity and the phase origin is arbitrary. Right: Corresponding power spectra in the range 0-1.5\,Hz, using Leahy normalisation (see text). The 90\% significance level for a mono-frequency signal is at a power value of 17 for each filter. 
       }
         \label{Figindiv}
   \end{figure*}


   \begin{figure*}
   \centering
  \includegraphics*[width=4.57cm,angle=-0,trim=50 60 90 40]{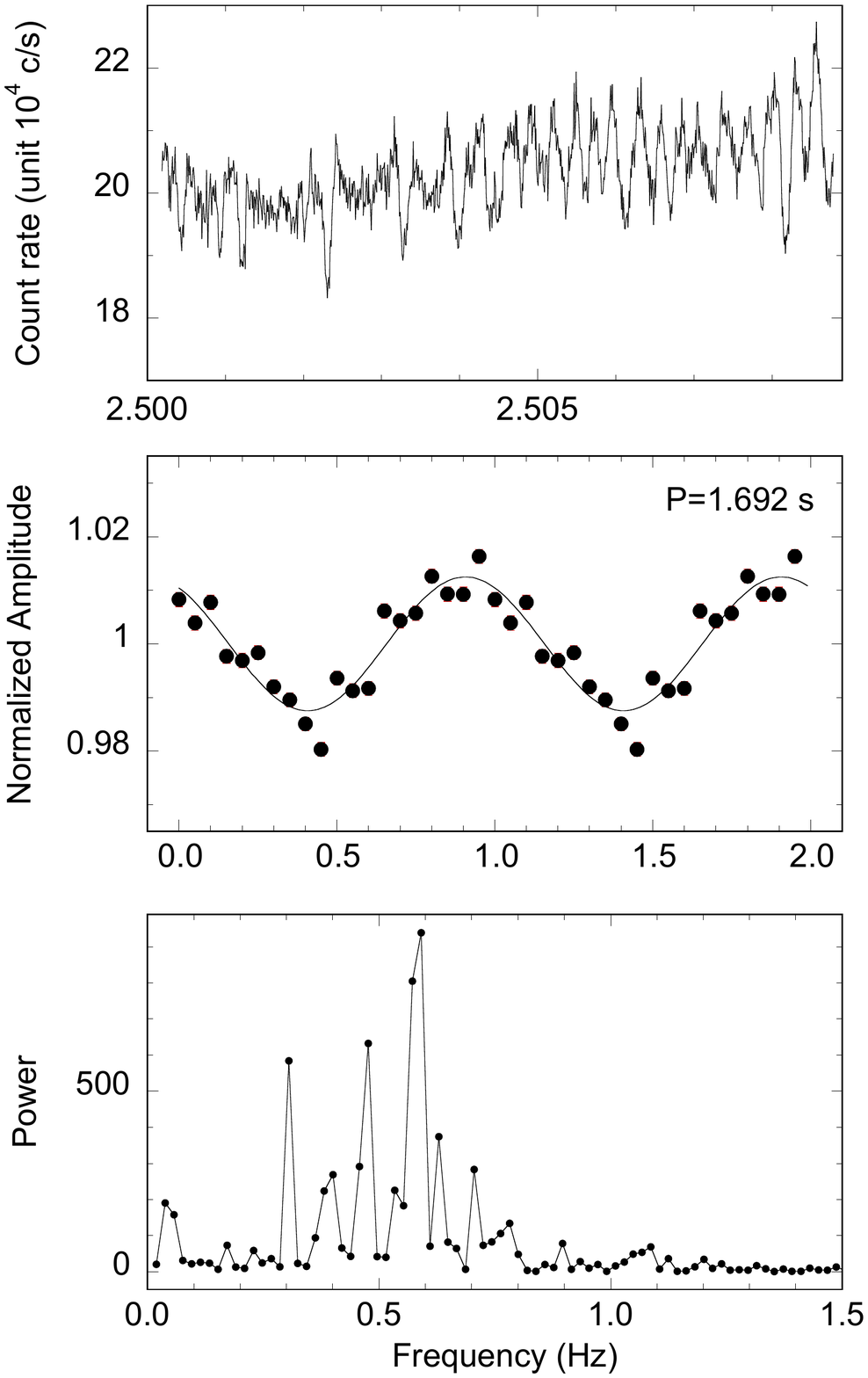} 
   \includegraphics*[width=4.4cm,angle=-0,trim=80 60 80 30]{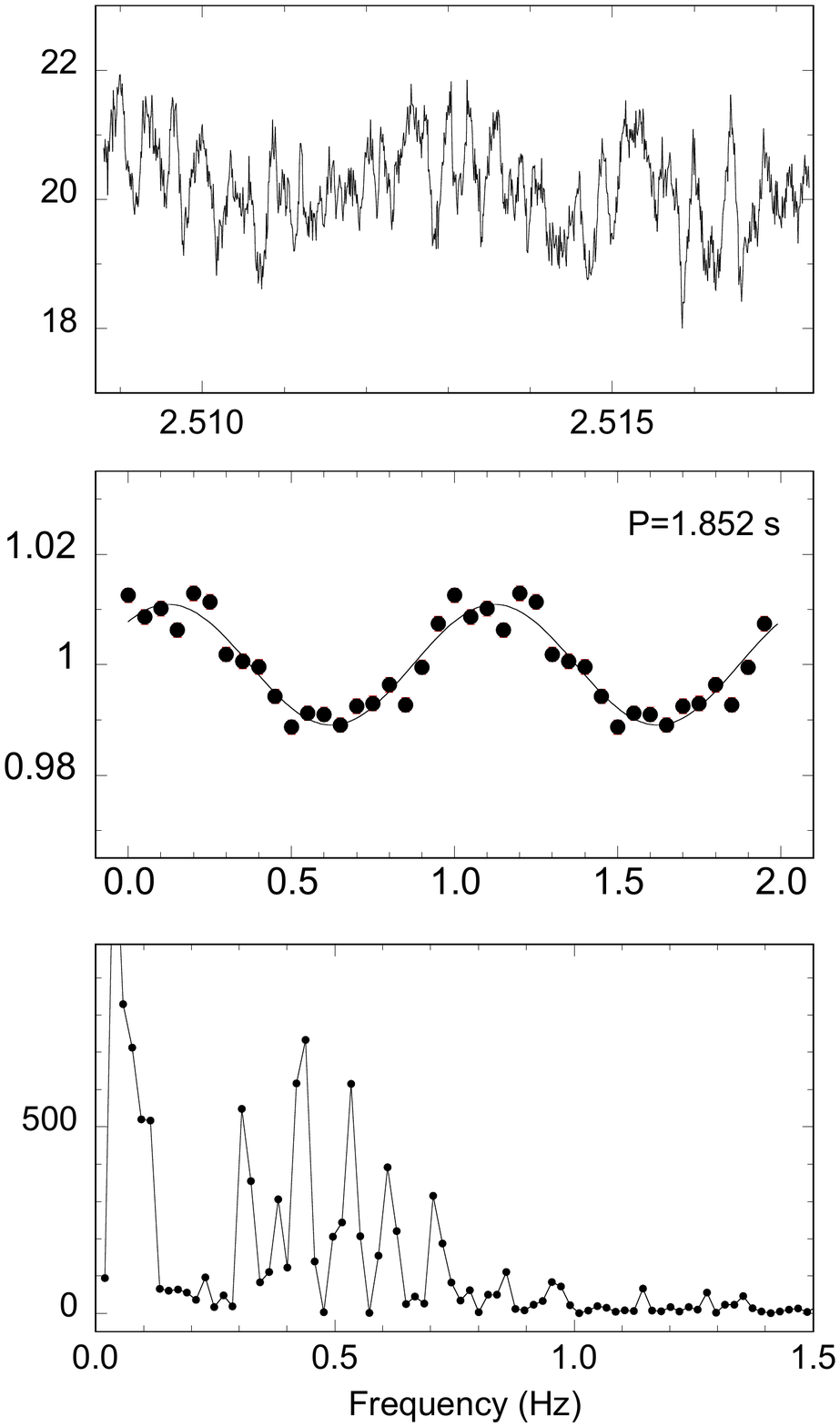} 
    \includegraphics*[width=4.4cm,angle=-0,trim=80 60 80 30]{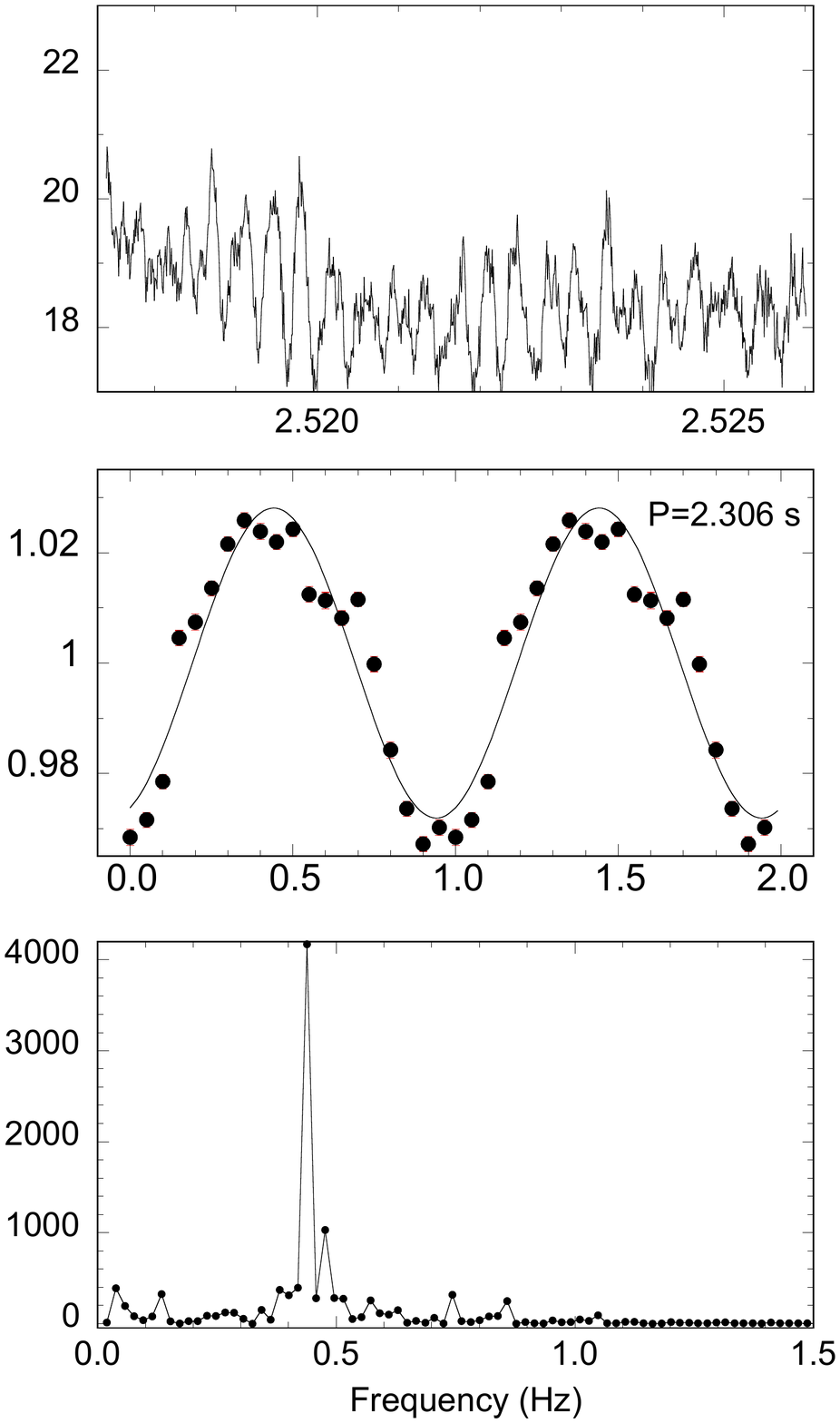} 
     \includegraphics*[width=4.4cm,angle=-0,trim=90 60 70 30]{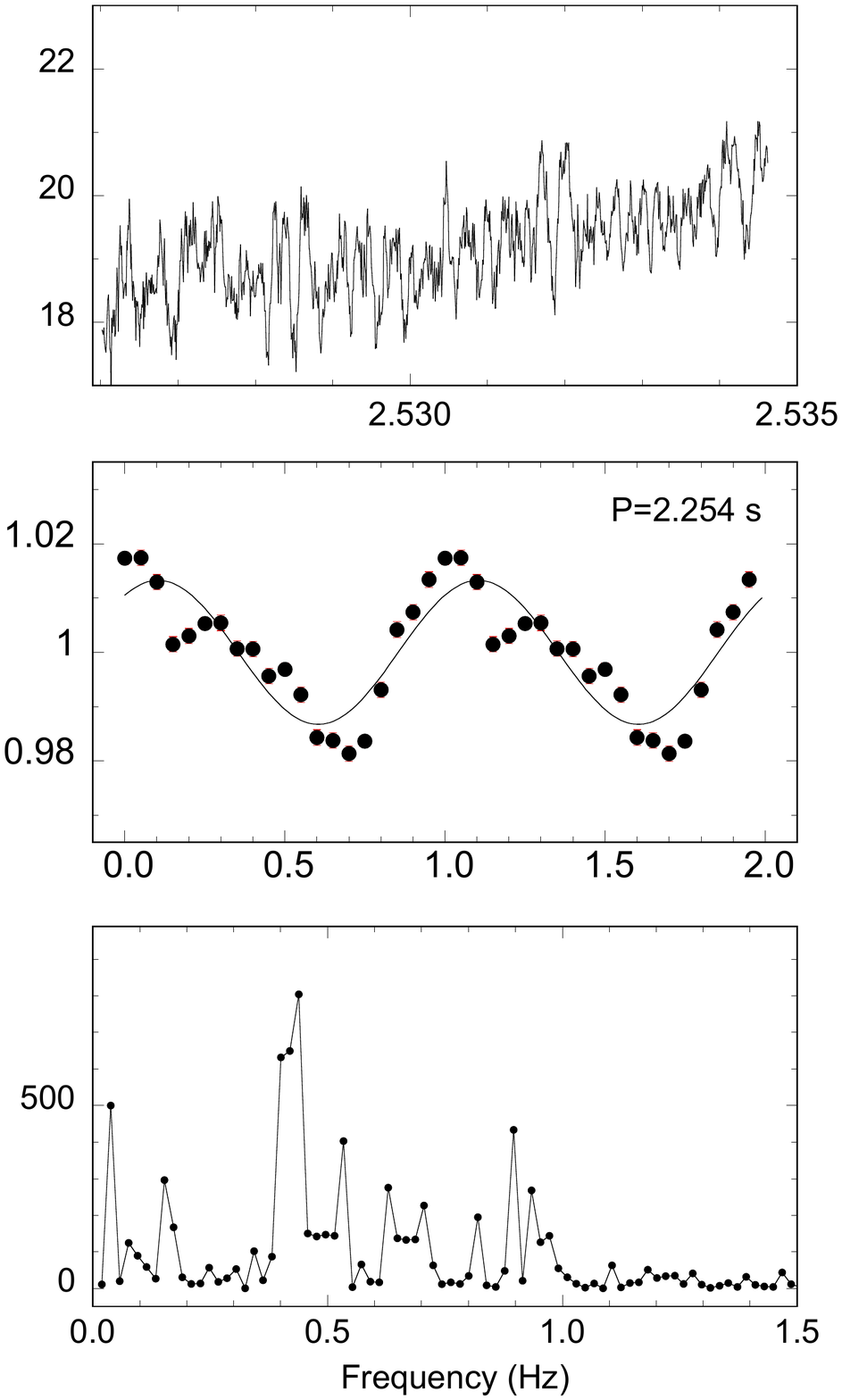} 
       \caption{Top: r' light curve of four contiguous 52.4\,s data segments with no overlap (covering the orbital phase interval 2.5 to 2.535 with no interruption). Middle:  Mean pulse shape. The QPO light curves are folded at the  best sinusoidal period indicated and repeated twice for clarity. Bottom: Corresponding power spectra using Leahy normalisation. The 90\% significance level is at a power value of 17. The Y-scale for power spectra  is the same for all segments except the third. 
      }
         \label{Fig4lcpsp}
   \end{figure*}

\subsection{QPO trains of oscillations}  \label{Trains}
The high quality of the ULTRACAM data obtained at VLT allows us to isolate data segments where flux oscillations are directly visible in the light curve. This allows a detailed study of the QPO frequencies as well as the pulse shape of the oscillations that could, in principle, be compared to numerical simulations.  

For illustration, we show in Fig.\,\ref{Figindiv},  a 52.4\,s segment of the light curves for the three filters obtained at an orbital phase  $\phi=1.273$.
This part is chosen as having the most significant power in r' and in He. 
Clear individual oscillations are seen in the r' and He light curves, but are much less visible in  u'.
When superimposed, the r' and He light curves are similar in shape, and cross-correlation shows no significant time lag.
The associated power spectra are also shown in Fig. \ref{Figindiv}. To evaluate the significance, the power is normalised here  according to
 \citetads{1983ApJ...266..160L}, 
 with  P$_{i}$ = 2\,$(a_{i})^{2}$/N$_{\gamma}$,   where  $a_{i}$ is the Fourier amplitude at frequency $\nu_i$ and N$_{\gamma}$ the total counts.
 With this normalisation, the power is independent of the source counts and the mean value of a Poissonian noise is equal to  2. 
All spectra show a similar complex shape with multiple significant peaks, indicating a superposition of independent oscillations. 
The r' and He filters shows the same main peaks at frequencies 0.40 and 0.48 Hz, while peaks in the  u' filter are at 0.48 and 0.53 Hz. 
We note, however, that the u' peaks are only marginally significant. 
With the adopted Leahy normalisation, the 90\% confidence level for a pure sinusoidal (mono-frequency) signal is at a power level of 17.

To derive a mean pulse shape, the data were folded at the mean best fit r' QPO period (2.556\,s) determined from a standard $\chi^{2}$ test.
 The light curves folded with this mean period and normalised by their mean values are shown in Fig. \ref{Figindiv}.  
 They show an almost sinusoidal shape and the best sinusoidal fit for each filter is displayed as a thin line.
In this interval, the computed amplitudes  for r', He, and u' are respectively  $4.30\pm0.26\,\%$, $ 4.25 \pm0.45 \,\%$, and$1.13 \pm 0.64\,\%$ (error bars at 90\% confidence level).


   \begin{figure*}
\includegraphics*[width=18.2cm,height=10.9cm,angle=-0,trim=160 90 380 80]{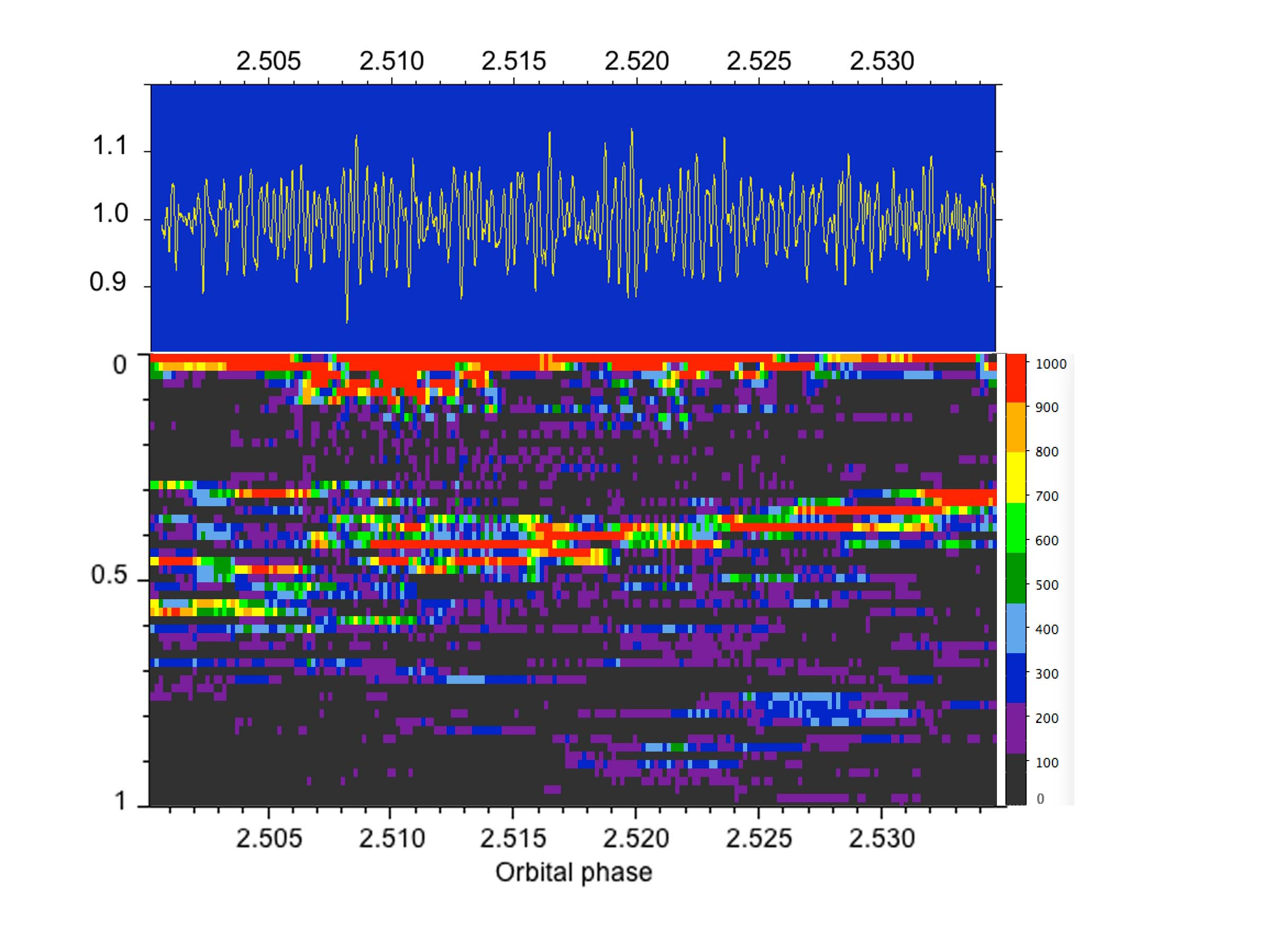} 
       \caption{
       Bottom: Two-dimensional time-frequency image of the rms fractional amplitude in a 210\,s time interval for the r' filter (same interval as in  Fig. \ref{Fig4lcpsp}). The X-axis is the orbital phase and the Y-axis is the frequency in the restricted range 0-1\,Hz, increasing from top to bottom. Colour-coded rms amplitude is shown according to the scale at right (in units of $10^{-6}$ rms$^{2}$.Hz$^{-1}$). Sliding 52.4\,s FFTs have been computed with a 1.28\,s step. 
       Top: Corresponding fractional r' light curve  shown after normalisation by dividing the original data by a 5\,s moving average and after  a 0.5\,s smoothing to better show the shape of oscillations.
       }
         \label{Fig4psp}
   \end{figure*}

Closer inspection of the 52.4\,s r'  light curve reveals that the individual $\sim$ 2.5 s oscillations are not regular, either in shape or in amplitude and duration.  
For instance, the two first oscillations in Fig. \ref{Figindiv} are strong, but the third  does not drop to its minimum, and subsequent pulses are not well defined. 
In addition, the two first peaks are separated by a shorter time than the average best period of 2.556\,s (0.391\,Hz). 
A shift is also seen at orbital phase 1.2744 and a pulse is essentially missing at phase 1.2757.

Significant variations are also often present from one data segment to the next and there are some
cases where narrow peaks appear or disappear on timescales of the order of the segment duration. As an example, Fig. \ref{Fig4lcpsp}  shows the light curves and power spectra for the r' filter in four consecutive 52.4\,s data intervals with no overlap. Complex trains of oscillations are seen in the first two intervals followed by a highly coherent single-frequency oscillation with large amplitude in the third segment and then degraded again in a more complex superposition in the last interval. All major peaks in the power spectra are well above the 99\% detection level  at a power value of 22. This type of behaviour occurs consistently throughout  the observation.

Figure \ref{Fig4psp} shows the 2D time-frequency image corresponding to this $\sim$ 210 s interval. To better show the temporal evolution, sliding 52.4\,s FFTs have been computed with a 1.28 s step, resulting in a 97\% overlap. The evolution of the power spectrum is clearly seen with complex multi-peaks of equivalent strengths spread over a wide range of frequencies and progressively evolving into a narrower frequency range as time progresses. Some significant power is also seen close to  twice the main frequency in the last part of the interval.
At the top  the oscillations in the source flux are also shown, normalised by dividing by a 5\,s moving average  and applying  a 0.5\,s smoothing. They display important variations in shape and amplitude.   


\section{Discussion}  \label{Discussion}

Quasi-periodic oscillations were detected throughout our $\sim$ 5.6\,h observation covering 3.3 orbital cycles.  Previously, oscillations were also nearly constantly observed during the source high state  
 (\citeads{1985A&A...145L...1L},   
  \citeads{1991ApJ...382..315M}, 
  \citeads{1992A&A...265..133L}).   
 The range of QPO frequency (0.3-0.8 Hz) is always similar, while the QPO amplitudes, expressed in terms of fraction of the total flux,  vary somewhat from date to date and also with wavelength. Our observations are the first to provide simultaneous QPO measurements at three different optical wavelengths, with high signal-to-noise ratio, and therefore provide interesting constraints on the energy distribution of the oscillations.
 
\subsection{Energy distribution of the oscillations}

According to the shock instability model, optical QPOs originate from the cyclotron emission of the accretion column and should scale with this emission. 
However,  contrasted with the hard X-rays range where the bremsstrahlung emission from the column is the dominant process,  in polar systems the optical cyclotron flux may be strongly diluted by other contributions. For example, the WD heated polar cap, the accretion stream, or the secondary star also radiate  in the optical. The true QPO pulsed fraction should therefore  refer  only to the cyclotron flux  and not to the total flux. As our observations are not flux calibrated, to determine the absolute energy spectrum of the QPOs we first make the reasonable assumption that the source spectrum during our observation is similar to that observed in an equivalent high state by 
 \citetads{1991ApJ...382..315M}.  
For this conversion, we make the approximation that our two u' and r' filters are not significantly different from the U and Rc in 
 \citetads{1991ApJ...382..315M}, 
 and indeed the estimated mean magnitudes for r'  (15.0) and u' (15.8)  (see Sect. 2) are in agreement with their U and Rc light curves.
The B filter, centred on 4400\,\AA , is also close to the central wavelength of  our He filter (4662\,\AA).

The non-cyclotron contribution to the optical flux was then evaluated from the spectrum reported for the source in low state by
\citetads{1990MNRAS.244P..20P},  
including mainly the contribution of the two stars.
The difference between the high- and low-state spectra should give a reasonable estimate of the accretion column cyclotron emission alone. 
By subtracting the non-cyclotron flux, the fractional QPO amplitude in the different filters is then revised as 2.58\%, 2.00\%, and 0.73\% respectively for the r', He, and u' filters. 
In the context of shock radiative instabilities, the QPO amplitudes are expected to be a constant fraction of the cyclotron flux at each wavelength, a result that differs significantly from the observed amplitudes. 
However, in the narrow He filter, the flux is also affected by the contribution of the broad \ion{He}{ii} $\lambda \, 4686$ emission line. A value of 37\,\AA{ } for the equivalent width of the line has been reported during high states 
\citepads{1987MNRAS.224..987R}.  
As the filter width is about 110\,\AA, the line contribution to the total flux in the filter is then $\sim$25\%. If the continuum alone is oscillating, the relative amplitude of QPOs must therefore  be corrected by a factor (1/0.75), leading to a corrected value of $\sim$2.68\%. Taking into account the uncertainty on the exact value of the line equivalent width at the time of the observation, this value appears sufficiently close to the r' value so that only the continuum is probably responsible for the oscillations. Contrary to what is suspected for AN UMa 
\citepads{1996A&A...306..199B}, 
the \ion{He}{ii} 4686\,\AA\, emission line does not appear here to contribute appreciably.

\begin{figure}
   \centering
 \includegraphics*[width=8.9cm,angle=-0,trim=70 210 90 260]{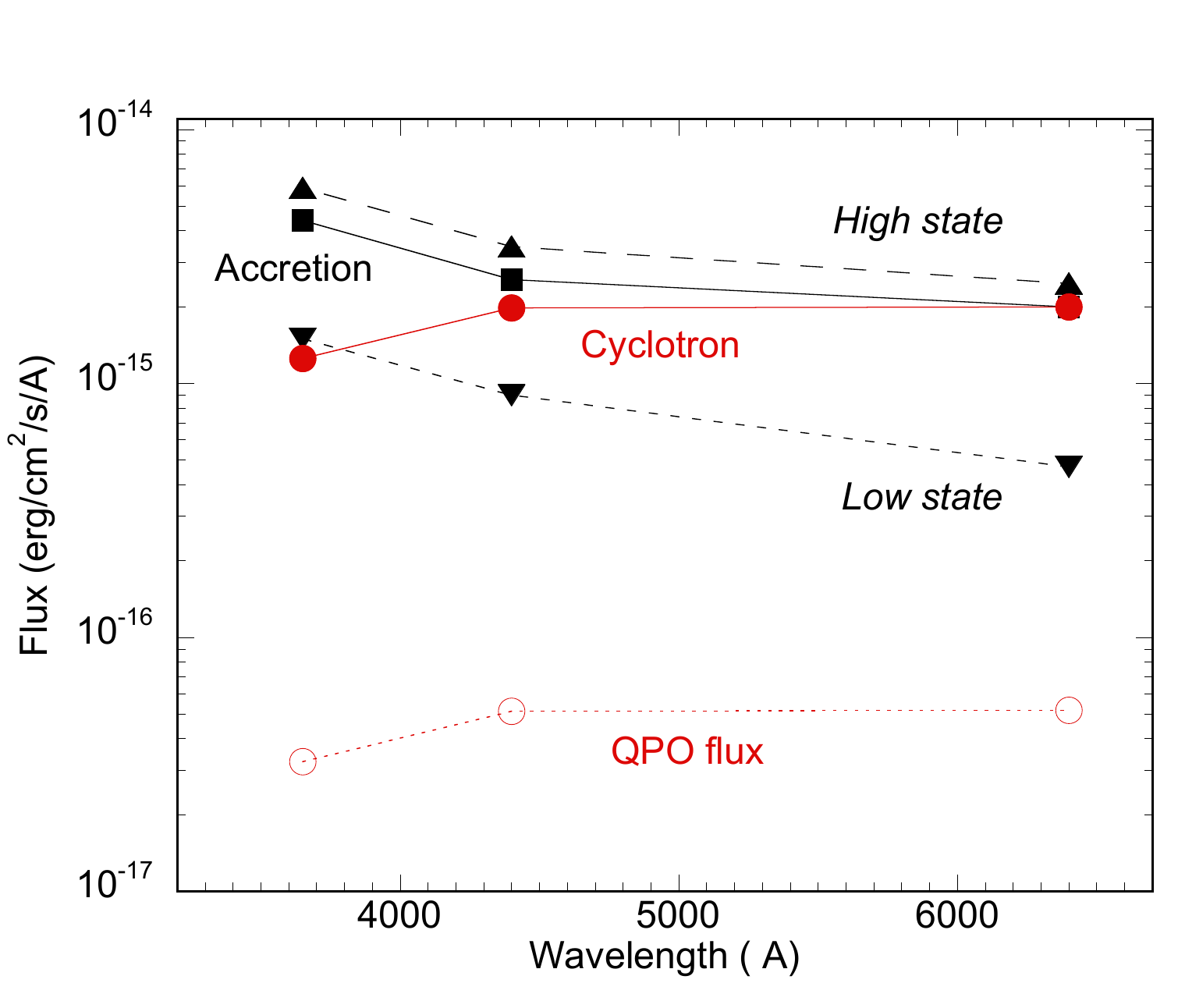}
       \caption{
      Time-averaged spectral distribution of the QPO and continuum flux  for V834 Cen. The high-state level (up-pointing  triangles) is from 
      \citetads{1991ApJ...382..315M} 
      and low-state level (down-pointing  triangles) from 
      \citetads{1990MNRAS.244P..20P}. 
      The difference (high-low) state (black squares) is assumed to be the accretion flux.  Also shown are the measured QPO flux (open red circles) and the cyclotron flux (filled red circles) reconstructed from the QPO amplitude, assuming a constant fraction of the cyclotron flux as expected from the models (see text). Note that there is a large difference in the u' filter when compared to the accretion flux.
       }
         \label{Figcyc}
   \end{figure}

QPO amplitudes are thought to be a direct measure of the pure cyclotron energy distribution. Figure \ref{Figcyc} shows the QPO energy distribution and also the expected accretion flux defined as the difference (high-low) state. Also shown is the reconstructed cyclotron emission assuming a constant fraction of 2.58\% as observed in the r' filter. The reconstructed distribution is roughly compatible with the expected cyclotron emission for the He filter, but a large discrepancy exists in the u' filter. This difference can be explained if, 
in addition to
the column, 
other sources of light also contribute at this short wavelength,  as noted earlier 
\citepads{1991ApJ...374..744S}. 
Among the possible contaminations is the heated surface of the white dwarf surrounding the base of the column. 
Assuming a 18 eV blackbody spot as derived from EUVE high-state observations 
\citepads{2002ApJ...578..439M}, 
the additional flux will require a spot with a fractional emitting area $f  \sim 0.02$ with respect to the white dwarf surface. 
This is significantly larger than the typical value $f  \sim 0.001$ deduced from the EUVE observations and may indicate more dilution in this band by other contributions such as the  Balmer discontinuity in emission or the heated hemisphere of the companion.


\subsection{Oscillations from simulated data}

\begin{figure}
   \centering
 \includegraphics*[width=8.9cm,angle=-0,trim=70 80 80 20]{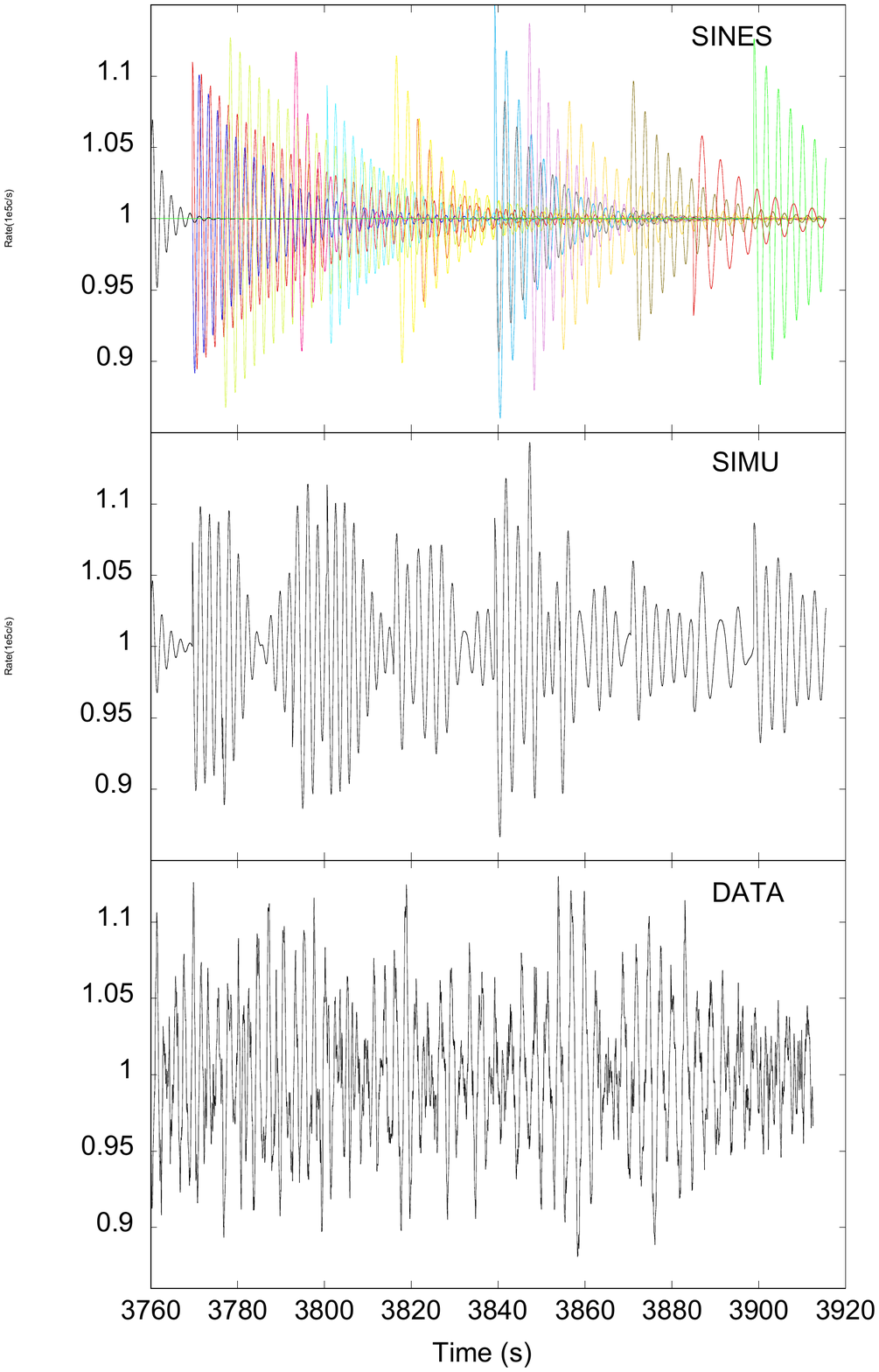}
       \caption{
      Reconstructed simulated data from superposed damped sinusoids. Top: Individual sinusoids with random start, amplitude, frequency, and damping time. Middle: Simulated data as the sum of damped sinusoids. Bottom: Observed data,  normalised by dividing by a 5\,s moving average,  at orbital phase ($\phi =1.08-1.11$) where a significant large QPO amplitude is observed. Similar features can be recognised in the data and in simulated light curves.  
       }
         \label{Figsimu}
   \end{figure}

The shape of the oscillations and the characteristics of the power spectra, most often showing multiple peaks even on short timescales, are suggestive of a superposition of modes. We attempted to reproduce the trains of pulses by performing simulations of light curves. We chose to add damped sinusoids with frequencies, amplitudes, phases, and damping time chosen randomly from Gaussian distributions around values and widths typical for QPOs. The start times of the sinusoids were also randomly chosen.  
We compare these simulations with selected portions of the smoothed r' light curve  obtained after  dividing the original data with a 5\,s moving average  and applying a 0.5\,s smooth to the result. Figure \ref{Figsimu} shows a typical simulation in a 157\,s interval (3x1024 points). Details of the 15 input sinusoids are shown (top), as well as the resulting light curve (middle) compared to a data interval (bottom) where the most significant oscillations are seen, with amplitude reaching $\sim$ 10\% (at orbital phase $\phi =1.08-1.11$). Similar features are visible in both simulation and data with a strong modulation of the QPO amplitude. The resulting mean power spectra for data and simulations are also structured with several similar  peaks in the same range of frequencies. Though this reconstruction is rather arbitrary and by no means unique, it shows that the observed trains of pulses can result from the superposition of individual damped pulsations.  

With such behaviour, the notion of pulse coherency is hard to define. If the coherency is defined as the ratio 
$\nu/\delta \nu$,  where   $\delta \nu$    
 is the peak width in the power spectra, then it is clear that on long timescales the coherency is always very low, with a value close to $\sim$2 derived from the mean QPO characteristics (see Table~\ref{table:1}). Even on shorter time intervals, narrow peaks are only seen to persist for not much more than one of our typical $\sim$ 52.4\,s intervals, leading to a $\nu$/$\delta \nu$ maximum coherency of $\sim$ 25. 

In this respect, we note that the description given in
\citetads{1992A&A...265..133L}   
is somewhat misleading.  
\citetads{1992A&A...265..133L}   
chose to define coherency as a ratio $P/|\dot{P}|$ where $\dot{P}$ is supposed to result from a regular frequency shift with time. With this definition, 
\citetads{1992A&A...265..133L}   
determined  a coherency of 
$P/<|\dot{P}|>$  $\approx$ 620\,s.  
We do not observe any regular systematic frequency drift in our data, nor could we define a significant $\dot{P}$ value. In fact, 
\citetads{1992A&A...265..133L}   
also mentioned that variations in $\dot{P}$ were erratic resulting in an average value of zero.
The coherency defined as $\nu/\delta \nu$  seems to be a much more reliable measure, and it indicates that pulses are erratically variable in amplitude and frequency on a typical maximum timescale of $\sim$ 20\,s. Inspection of Fig. \ref{Figsimu} indeed reveals that trains of pulses have typical duration of (10-20) s or (5-10) cycles.


\subsection{Comparison with RAMSES numerical simulations}
From previous simulations of the accretion shock instability based on a hydrodynamical code HADES 
(\citeads{2011Ap&SS.336..175M}),  
amplitudes of the expected oscillations in X-rays and optical have been derived and compared to observations for a large set of polars 
(\citeads{2015A&A...579A..24B},  
\citeads{2015A&A...579A..25B}). 
 In an effort to improve the simulations with a planned future 3D treatment, we have adapted the RAMSES code 
(\citeads{2002A&A...385..337T})  
  to simulate the radiative accretion dynamics in the accretion column (Van Box Som et al. 2017, in preparation).
 This multidimensional hydrodynamic Adaptive Mesh Refinement (AMR)  code has been modified to solve the radiation hydrodynamics 
 equations on a 1D Cartesian and uniform grid.  The resolution of the grid cells has been optimised to have an accuracy of about $0.1$ km for each simulation. Radiative losses are added as a source term in the energy equation and the effect of the WD gravity  is taken into account.
 At the upper boundary before the shock, Dirichlet conditions are used to model the homogeneous incident flow from the low-mass companion star. They are characterised by the WD mass which determines the free-fall velocity and by an accretion rate which defines the density ($\rho_{0}$)  of the incoming flow. Since the Mach number of the incoming flow is around 30, a Harten-Lax-van Leer Central wave (HLLC) scheme is implemented in RAMSES to capture shocks and solve high Mach number flows.\\
At the bottom of the column near the white dwarf surface, a specific boundary condition has been implemented in the boundary region to mimic the accretion onto the photosphere of the white dwarf, modelled by 
\begin{equation}\label{eq}
[\rho V (t)]_{WD} = [\rho V(t)]_{F} \left[1-2\left(\frac{\rho_{0}}{[\rho(t)]_{F}}\right)\right]
,\end{equation}
where $[\rho V(t)]_{WD}$ is the mass flux in the WD photosphere, and $[\rho V(t)]_{F}$ and $[\rho(t)]_{F}$ are respectively the mass flux and the density of the incident flow just above the WD photosphere in the accretion column. The mass and the energy are conserved between the column and the photosphere. At the first iteration of time, the WD photosphere behaves as a perfect wall because
$[\rho]_{F}=\rho_{0}$ and then $[\rho V]_{WD} = -[\rho V]_{F}$. 
Since radiative losses increase the accreted matter near the WD photosphere, this later gradually  absorbs the accreted mass, progressively becoming more porous. Thus, the system is locally self-regulated.
Finally, the radiative losses in the accretion column are modelled by a cooling function $\Lambda(\rho, T)$, expressed as the sum of the bremsstrahlung and the cyclotron processes. The optically thin bremsstrahlung and the optically thick cyclotron cooling are respectively exactly and approximately defined by a power-law function
(\citeads{1994ApJ...426..664W}). 

The code was run for a set of parameters typical for the source with a WD mass of M$_{WD}$=$0.66$ M$_{\odot}$, a magnetic field of B = 23 MG, and a total accretion rate of $\dot{M}$ = $0.14 \times 10^{16}$ g.s$^{-1}$, corresponding to a total X-ray luminosity of L$\rm _x$= $1.5 \times 10^{32}$ erg.s$^{-1}$ (see 
\citeads{2015A&A...579A..24B}). 
Computations were performed at 3000 to 5000 steps, with time resolution adapted to ensure  coverage of  at least 160 cycles of oscillations. To see the effect of the WD mass, simulations were also done for M$_{WD}$=$0.85$ M$_{\odot}$ 
 (\citeads{2004MNRAS.348..316P}), 
and the simulations were performed varying the unconstrained column cross section S in the range ($10^{12}$-$10^{16}$)\,cm$^{2}$.
In each case, typical outputs were produced including the X-ray (0.5-10\,keV) and cyclotron light curves, computed using the local bremsstrahlung and cyclotron emissivities
combined with the density and temperature profiles extracted from the simulations.  The amplitudes of the oscillations in X-rays (bremsstrahlung) and optical (cyclotron) were recovered from the associated FFTs. 
When the oscillations show power distributed at different frequencies, the total quadratic sum amplitude was computed (see 
\citeads{2015A&A...579A..25B} 
for details). 


   \begin{figure}
   \centering
 \includegraphics*[width=8.9cm,angle=-0,trim=30 0 250 350]{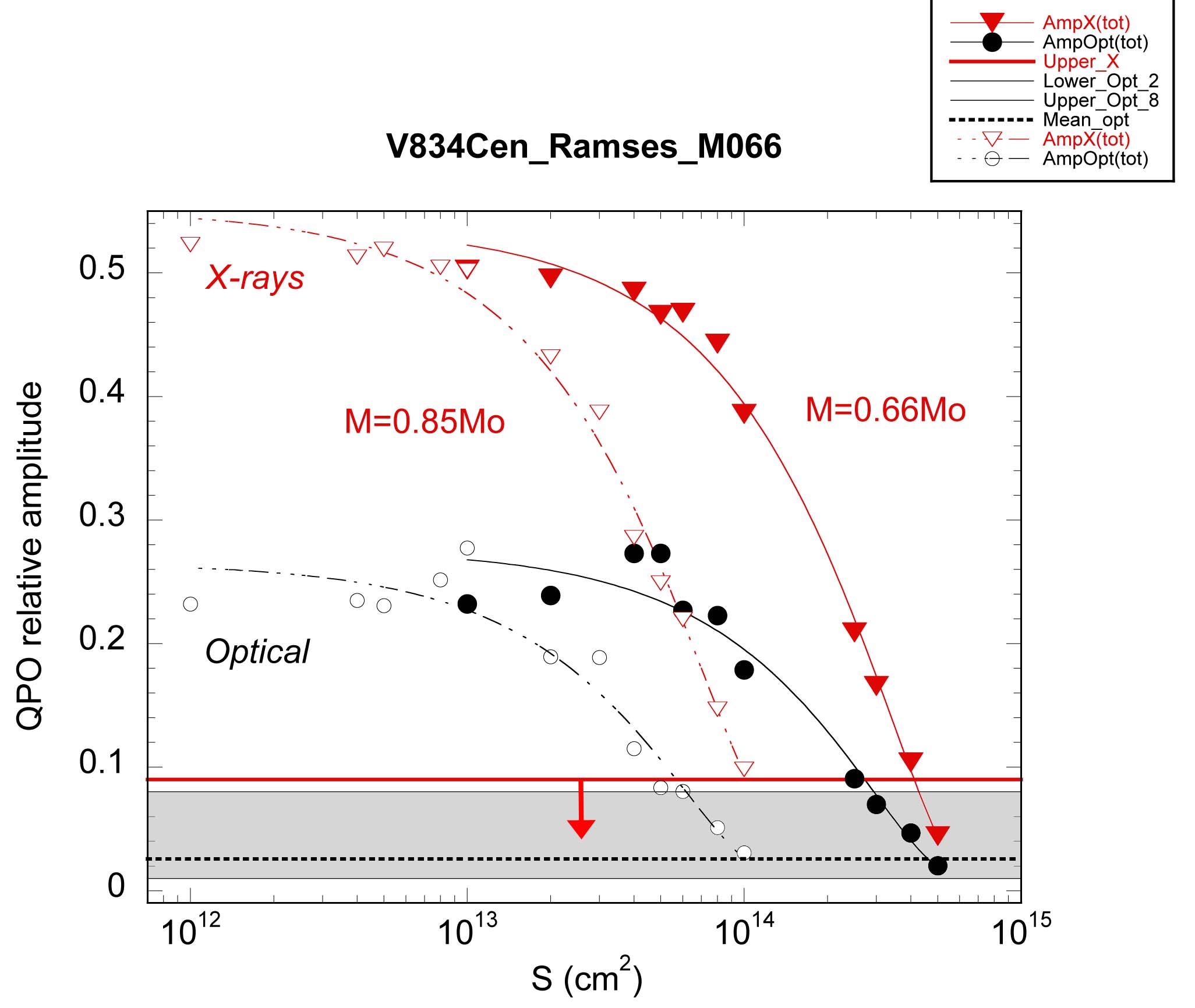}
       \caption{
      QPO amplitudes from RAMSES numerical simulations. QPO relative amplitudes in X-ray (0.5-10\,keV) bremsstrahlung (in red) and optical cyclotron (in black)  are shown for two different masses M$_{WD}$=$0.66$ M$_{\odot}$  (filled symbols) and $0.85$ M$_{\odot}$ (open symbols) as a function of the column cross section. Full and dotted lines are polynomial fits to the data points. Also shown as horizontal lines are the observed upper limit in X-rays (red line) and the mean optical amplitude (dotted line) and range (shaded area) from the present observations. 
       }
         \label{Figramses}
   \end{figure}

Figure \ref{Figramses} shows the QPO predicted amplitudes in X-rays (bremsstrahlung) and optical (cyclotron) as a function of the column cross section  for the two masses M$_{WD}$=$0.66$ and 0.85 M$_{\odot}$. Typical large $\sim 45 \%$ and $\sim 25 \%$ amplitudes are expected, respectively in X-rays and optical, for a range of  cross section of ($10^{13}$ - $10^{14}$)\, cm$^{2}$. For a mass of $0.66$ M$_{\odot}$, this corresponds to a rather large specific accretion rate (140 - 14) g.cm$^{-2}.$s$^{-1}$, ensuring that the bremsstrahlung cooling dominates over the cyclotron. Also shown in  Fig.\,\ref{Figramses} is the upper limit on QPO X-ray amplitude (9\%) derived from XMM-Newton observations
(\citeads{2015A&A...579A..24B}), 
as well as the mean value corrected for dilution (2.6\%) and range (shaded area) of QPO optical amplitudes from these observations. The observed values are compatible with the simulations only in a narrow range of cross sections around $\sim$ (4-5) $\times 10^{14}$\, cm$^{2}$. This is comparable with the more approximate values derived from mass and cross section interpolations given in 
\citetads{2015A&A...579A..24B}. 
As expected, increasing the mass to $0.85$ M$_{\odot}$ shifts the possible values to lower cross section since, due to higher gravity, the same X-ray luminosity is achieved for a lower total accretion rate and therefore the same specific accretion rate is only reached for lower cross section. We note that for this mass, no  detectable oscillations are produced over S > $10^{14}$\, cm$^{2}$.

As noted earlier (e.g. 
\citeads{1991ApJ...378..665I},  
\citeads{2015A&A...579A..24B}),  
there is, however, no agreement with the frequencies predicted  by the simulations. For a M$_{WD}$ = $0.66$ M$_{\odot}$ and S = (4-5)$\times 10^{14}$\, cm$^{2}$, the QPO optical frequencies are found at 
$\nu$ = (14-19) Hz (P=0.05-0.07\,s), largely at variance with the observed range (0.25-5) Hz. Except for a very small secondary excess at  
$\nu$ $\sim$ 12 Hz, there is no significant signal at lower frequency in the simulations. Detailed results of RAMSES simulations, both in terms of QPO amplitudes and frequencies, will be discussed elsewhere (van Box Som et al. 2017, in preparation). 

This discrepancy is the major problem in interpreting the oscillations in terms of the shock oscillation models. 
As there are presently no significant limits from QPO observations at these high frequencies, either in X-rays or optical, there is still a possibility that these fundamental frequencies are present but remain undetected. 
However, the origin of the $\sim$ 0.5 Hz oscillations remains unsolved. 
In this respect, 1D simulations of a homogeneous column are probably too crude an approximation to provide relevant results.
In fact, though the accreted matter is thought to be captured in the orbital plane near the inner Lagrangian point (L$_1$), small instabilities near this region may lead to the capture of a flow with variable density along different field lines generating an inhomogeneous column or the presence of independent magnetic tubes with variable densities. 
Such a configuration has also been suggested from 2D magneto-hydrodynamical simulations in the context of radiative accretion shocks above young stellar objects (YSOs) 
(\citeads{2013A&A...557A..69M}). 
For low-$\beta$ (dynamic/magnetic pressure) plasma, the presence of `fibrils', flux tubes that can oscillate independently, has been demonstrated. 
This approach is particularly promising in the case of Polars where the $\beta$ value is at least two orders of magnitude lower than for YSOs, due to high magnetic fields. 
It probably requires a full 3D treatment to deal with possible transverse effects in the column across the different tubes. 
In particular, frequencies lower than that derived from the typical radiative cooling time could possibly arise from constructive instabilities across fibrils with different mean densities and/or cross sections. 
A significant coupling between the different tubes could, for instance, occur through processes like transverse Alfven waves, leading to beat frequencies in the observed lower range.

\section{Conclusion}
High-quality data with good S/N obtained at the VLT using the ULTRACAM camera allow an unprecedented study of the QPO variability in the polar V834 Cen. These first simultaneous QPO measurements at three different optical wavelengths over more than three consecutive orbital cycles of the source provide several important conclusions:

- The QPOs are always present during this high-state observation
 and when the S/N is sufficient, as it is for the r' filter, the QPOs are found over the full orbital cycle, including at minimum flux. This is at variance with previous results where no significant QPO amplitude could be measured 
 during orbital minimum,  but these results were based on less sensitive data 
(\citeads{1992A&A...265..133L},   
\citeads{1991ApJ...382..315M}). 
According to the source characteristics (orbital inclination and magnetic colatitude), the column is never significantly eclipsed at this phase, consistent with a cyclotron origin
close to the white dwarf. 

- QPOs are seen in the narrow He filter, centred close to
the \ion{He}{ii} 4686\,\AA\, line, but their amplitudes are comparable to those expected from the underlying continuum only and, unless there is an additional important dilution effect, the emission line does not contribute to the oscillations.

- During the orbital bright  phase,
there is a clear correlation of the QPO absolute amplitude with the flux over the orbital cycle so that the relative QPO amplitude remains approximately constant at a mean level of $\sim$2.4\% (for the r' filter). 
However, some additional variability is  present in the orbital bright phase with isolated peaks up to $\sim$5\% that are not obviously related to the source flaring activity. 
Through the orbital faint phase, the relative amplitude drops to a lower value of $\sim$1\%. 
This may be explained by a significant dilution of the cyclotron flux at this phase, due to additional contributions such as the WD heated surface and/or the illuminated secondary or accretion stream.

- The energy distribution of the oscillation flux  is comparable with that expected from a cyclotron origin for the r' and He filters, but requires significant dilution effects in the u'  filter where it is found to be less than expected.

- On timescales of the order of $\sim$(10-20 s), the QPOs have a rich and complex variability.
Though the mean amplitude on timescales longer than  $\sim$ 1 minute  remains around 2-3\%,
oscillations of higher amplitudes up to $\sim$10\% are seen over short intervals (10-20 s)  with a quasi-sinusoidal pulse shape. The typical coherency is therefore of the order of $\sim$ (5-10) cycles. 
The rapid frequency and amplitude variability reported in this paper clearly demonstrate that QPOs are produced by the superposition of simultaneous oscillations of different frequencies at a given time rather than by one unique oscillation with a variable frequency with time. This points to a more complex structure than a homogeneous column. 
Because the QPO frequency varies as $\dot m$ (to the first order, considering the bremsstrahlung term only), the overall change in frequency (of the order of $\sim 30\% $) can be accounted for by a similar $\dot m$ variation. This variation can be produced by individual blobs with different densities and/or different sections falling along different magnetic field lines.

- According to 1D numerical simulations of radiative shocks, for the set of parameters corresponding to V834 Cen, radiative instabilities can develop and produce QPOs of significant amplitudes for a column cross section S $\lesssim$ (4-5)$\times 10^{14}$\,cm$^{2}$. For a mass of M$_{WD}$=$0.66$ M$_{\odot}$ and S = 5$\times 10^{14}$\,cm$^{2}$, corresponding to a specific accretion rate of 2.8 g.cm$^{-2}.$s$^{-1}$, QPO amplitudes of 4.4\% and 2.0\% are expected respectively in X-rays and the optical, in agreement with the reported X-ray upper limit and optical values reported here. 
However, the QPO predicted frequency ($\sim$19 Hz) is largely inconsistent with the observed one. 
This is a major discrepancy with the shock instability model.
 In view of the general agreement on the association of the QPOs with the cyclotron emitting region, it is likely that cooling instabilities are at the origin of the QPOs but operate in a more complex way than via a homogeneous column with steady accretion rate. 
 Variable accretion through independent flow tubes has to be investigated further by 3D simulations to evaluate the contribution of possible lower frequency instabilities arising from interaction between the different tubes.
We note that the observed QPO power spectrum is presently restricted to a rather limited range ($\nu$ $\lesssim$10 Hz). 
This  could actually be  just the tip of the iceberg, masking a much more complex temporal structure where both low frequencies and a rich spectrum of high frequencies ($\nu$>10-500) Hz may be present, corresponding to individual tubes with slightly different regimes. 
For neutron star low-mass X-ray binaries (LMXBs), the wide range of   millisecond QPO spectra 
was only uncovered  long after the source discoveries 
(\citeads{ 2006csxs.book...39V}). 
Polars are considerably fainter than LMXBs in X-rays, so similar studies will require future high throughput - high time resolution X-ray missions, and even in the optical may have to wait for the next generation of  30 m telescopes to efficiently investigate the accretion shock.

\begin{acknowledgements}
We wish to thank Olga Alexandrova for her valuable assistance in the wavelet analysis. 
This paper is based on observations collected at the European Organisation for Astronomical 
Research in the Southern Hemisphere under ESO programme 075.D-0846.
VSD and ULTRACAM are supported by the STFC (grant ST/J001589/1).
\end{acknowledgements}

\bibliographystyle{aa}         
\bibliography{V834Cen_Ultracam}    

\end{document}